\documentclass[aps,pre,twocolumn,amsmath,amssymb,showpacs,amsfonts]{revtex4}
\usepackage{epsfig}
\usepackage{graphicx}
\usepackage{dcolumn}
\usepackage{bm}

\begin{document}
\title{Universal statistics of density of inertial particles sedimenting in turbulence}

\author{Itzhak Fouxon$^{1}$}\email{itzhak8@gmail.com}
\author{Yongnam Park$^2$}
\author{Roei Harduf$^3$} 
\author{Changhoon Lee$^{1,2}$}\email{clee@yonsei.ac.kr}

\affiliation{$^1$ Department of Computational Science and Engineering, Yonsei University, Seoul 120-749, South Korea}
\affiliation{$^2$ Department of Mechanical Engineering, Yonsei University, Seoul 120-749, South Korea}
\affiliation{$^3$ Raymond and Beverly Sackler School of Physics and Astronomy, Tel Aviv University, Ramat Aviv, Tel Aviv 69978, Israel}

\begin{abstract}

We solve the problem of spatial distribution of inertial particles that sediment in Navier-Stokes turbulence with small ratio $Fr$ of acceleration of fluid particles to acceleration of gravity $g$. The particles are driven by linear drag and have arbitrary inertia. We demonstrate that independently of the particles' size or density the particles distribute over fractal set with log-normal statistics determined completely by the Kaplan-Yorke dimension $D_{KY}$. When inertia is not small $D_{KY}$ is proportional to the ratio of integral of spectrum of turbulence multiplied by wave-number and $g$. This ratio is independent of properties of particles so that the particles concentrate on fractal with universal, particles-independent statistics. We find Lyapunov exponents and confirm predictions numerically. The considered case includes typical situation of water droplets in clouds.

\end{abstract}


\maketitle

\section{Introduction}

It is rare in the theory of turbulence that one can make quantitative predictions \cite{Frisch}. In this paper we manage to provide complete, quantitative predictions on spatial behavior of particles driven by the turbulent Navier-Stokes flow $\bm u(t, \bm x)$ according to the law 
\begin{equation}
\ddot {\bm x}(t) =-\left(\dot {\bm x}(t)-\bm u[t, \bm x(t)] \right)/\tau+ \bm g,
\end{equation}
where $\bm x(t)$ is the particle's coordinate, $\tau$ is the Stokes relaxation time and $\bm g$ is the gravitational acceleration. We demonstrate that when the ratio $Fr$ of typical acceleration of fluid particles to $g$ is much less than one then the particles distribute over multi-fractal structure in space. This prediction holds independently of properties of particles such as $\tau$. 

Further the statistics of the spatial concentration of particles is log-normal. The spectrum of fractal codimensions \cite{HP} is linear $D_{\alpha}=D_{KY}\alpha$ where $D_{KY}$ is the Kaplan-Yorke dimension \cite{KY}. In the particular case where $\tau$ is comparable with time-scales of turbulence one finds 
\begin{eqnarray}&&\!\!\!\!
D_{KY}=\frac{15\pi \int_0^{\infty}E(k)k dk}{32g}, \label{kapyork}
\end{eqnarray}
where $E(k)$ is the spectrum of turbulence. Thus $D_{KY}$ determines the (fractal) statistics of spatial distribution of particles completely. 
Its $\tau-$ independence signifies universality of spatial distribution of particles which becomes independent of particles' size or density in $Fr\ll 1$ limit. In particular the pair-correlation function of the particles' concentration $n$ (preferential concentration) is 
\begin{eqnarray}&&\!\!\!\!
\langle n(0)n(\bm r)\rangle=\langle n\rangle^2\left(\frac{\eta}{r}\right)^{2D_{KY}},\label{pref}
\end{eqnarray}
where angular brackets stand for spatial averaging and $\eta$ is the smallest, Kolmogorov scale of turbulence \cite{Frisch}. Though $D_{KY}\propto Fr$ is small the preferential concentration is arbitrarily large when $r\to 0$. 

Finally we describe completely the spectrum of the Lyapunov exponents $\lambda_i$ that provide the logarithmic rates of growth of the axes of the ellipsoid into which turbulence deforms small volumes of particles, 
\begin{eqnarray}&&\!\!\!\!
\frac{\lambda_1 \tau}{D_{KY}}=-\frac{\lambda_3 \tau}{D_{KY}}=\frac{32}{75},\ \ \lambda_2=-\frac{\lambda_1 D_{KY}}{3}. \label{spctr}
\end{eqnarray}
Thus $D_{KY}/[\lambda_1 \tau]$ is universal constant that is independent of properties of particles, turbulence or $g$. The largest axis of the ellipsoid grows ($\lambda_1>0$) but the smaller ones decrease ($\lambda_{2, 3}<0$) signifying that turbulence transforms small volumes of inertial particles into cigars (in contrast, small volumes of fluid particles for which $\lambda_{1, 2}>0$ are transformed into pancakes). We confirm the predictions numerically. 

We demonstrate the transition in $Fr$ from singular density with $\langle n^2\rangle= \infty$ like in Eq.~(\ref{pref}) to continuous one with finite 
$\langle n^2\rangle$. When the dimensionless Stokes number $St$ measuring the particle's inertia $\tau$ is fixed at $St\gtrsim 1$ the increase of $Fr$ from small values brings the transition at $Fr\sim 1$. This is quite similar to the transition in $St$ when gravity can be neglected \cite{Bec}. When $St\ll 1$ the particles distribution is singular independently of $Fr$.  

The described results provide significant progress to the well-studied area. The asymptotic independence of a fractal dimension on $St$ was found in numerical simulations in  \cite{Becgr} (see statement of independent work below), see also \cite{Gust}. When gravity is strong, a new pattern of vertical clustering of sedimenting particles was recently observed \cite{Park2014}. However, no predictions comparable to Eqs.~(\ref{kapyork})-(\ref{spctr}) that are confirmed numerically were known so far. This is possible thanks to the observation \cite{FHa} that when $Fr\ll 1$ the particle's velocity is determined uniquely by its position so that there is well-defined flow of particles $\bm v(t, \bm x)$. In contrast to the cases considered previously there is no explicit formula for $\bm v(t, \bm x)$ via $\bm u(t, \bm x)$. Despite this we manage to perform the study using recently found \cite{Fouxon2,Fouxon1} universality in the distribution of particles in weakly compressible flows, see below. 

Remarkably the case of $Fr\ll 1$ is rather typical for water droplets in warm clouds so that the described results are of direct practical relevance.  

Inhomogeneity (preferential concentration) in the steady state distribution of particles in space originates from the particle's drift outside of turbulent vortices \cite{Maxey}. This drift, however small it is, can have large impact on the spatial distribution of particles because it accumulates with time. This becomes clear by considering inertial particles in simple rotation flow in which inertia-less particles would circle forever but inertial ones drift to the flow region's boundary. Thus results of evolution of uniform initial distributions of inertial and inertialess particles are completely different: inertial particles will be on the boundary with zero volume, inertialess ones will stay uniformly distributed over all the volume. Similar physical mechanisms are at work in turbulence whose vortices can be considered as rotation regions with boundaries serving as centers of accumulation of particles. These mechanisms continue to apply when gravity is relevant to particles' motion. The crucial question is whether the finite life-time of turbulent vortices will destroy the accumulation of inhomogeneities that holds for simple rotation. 

Lots of research was devoted to the study of the posed question recently \cite{MaxeyRiley,
Maxey,BFF,FFS1,Bec,Falkovich,BGH,Collins,BecCenciniHillerbranddelta,Stefano,Cencini,Olla,MehligWilkinson,MW,Shaw,Seinfeld,Flagan,planetary1,
Engineering1,Engineering2,Biology1,Biology2,Fouxon1,Fouxon2,Becgr,Gust,Park2014,FFS,FHa,FP1,FP2,Nature2013,CYL,LYC,JYL,CKL,AL1,AL2}. It results from these studies that the well-established paradigm that the turbulent flow mixes small particles 
all over the available volume of the flow (see e. g. \cite{reviewp} and references therein), can only be applied with a reservation. When the resolution at which the particles' distribution in space is resolved is increased, it turns out that the flow distributes the particles over a random strange attractor - a multi-fractal set in space \cite{HP,Eckmann}. Instead of filling the whole volume point particles concentrate on set with zero volume quite like particles in simple rotation flow. This conclusion gets less paradoxical, when one observes that mixing presumes coarse-graining of the particles' density over an infinitesimal scale: otherwise the particles form an intricate set which is dense in space but is not the whole space, see e. g. \cite{Dorfman,Eckmann}. Inertia causes the coarse-graining scale over which the mixing holds to have a finite lower limit $l$. At scales smaller than $l$ the particles' density is extremely non-uniform, being singular for point particles. This is the result of accumulation in time of the repel of inertial particles by turbulent vortices which  produces a non-trivial spatial distribution  \cite{MaxeyRiley,
Maxey,BFF,FFS1,Bec,Falkovich,BGH,Collins,BecCenciniHillerbranddelta,Stefano,Cencini,Olla,MehligWilkinson,MW,Shaw,Seinfeld,Flagan,planetary1,
Engineering1,Engineering2,Biology1,Biology2,Fouxon1,Fouxon2,Becgr,Gust,Park2014,FFS,FHa,FP1,FP2,Nature2013,CYL,LYC,JYL,CKL,AL1,AL2}. Since real particles always have inertia, the effect has much practical importance for a wide variety of natural phenomena and industrial processes, such as rain formation in clouds \cite{Shaw, FFS1}, aerosols spread out in the atmosphere \cite{Seinfeld, Flagan}, planetary physics \cite{planetary1}, transport of materials by air or by liquids \cite{Engineering1}, liquid fuel combustion engines \cite{Engineering2}, plankton population dynamics \cite{Biology1, Biology2,Nature2013} and more. Thus in any concrete application one should
evaluate $l$ to decide whether the particles should be considered as being mixed by the flow, or rather transported to a non-trivial multi-fractal set in space.

Recently a complete description of the statistics of the particles' density at small particle's size was provided in \cite{Fouxon1,Fouxon2} by using considerations independent of any modeling of the turbulent flow and thus applying to the Navier-Stokes turbulence. The statistics of the density field is log-normal and it can be derived fully from the pair-correlation function. The analysis neglected gravity and relied on the existence of the velocity field of the particles that prescribes uniquely the velocity of the particle in terms of its position \cite{Maxey}. The field exists if
the dimensionless Stokes number, $St$, given by the ratio of the relaxation time of the particle's velocity to the smallest Kolmogorov time-scale of turbulence \cite{Frisch}, is small. Since $St$ is proportional to the square of the particle's size, the study applies to small particles. The influence of the remaining dimensionless parameter - the Reynolds number $Re$ is to strengthen the formation of the fractal so it gets pronounced at smaller Stokes numbers: fractal co-dimensions are proportional to the product of $St^2$ and a power of $Re$ with positive, albeit small, exponent $\Delta$, see \cite{Fouxon1} and references therein. Due to the smallness of $\Delta$ the dependence on $Re$ can be often disregarded. When $St>>1$ the particles move through the flow quasi-ballistically rapidly traversing the vortices. 

In this paper, we study the impact of gravity on the particles' preferential concentration in space. This is demanded to apply the theory to nature. Gravitational settling causes the particles to sample vortices differently producing a new kind of preferential concentration that only recently started to be studied, see \cite{Becgr,Gust,Park2014} and references therein. We determine the asymptotic range of $St$ and $Fr$ for which the particles form fractal in space. We observe that this range coincides asymptotically with the range of definability of the flow of particles where the particle's velocity is determined uniquely by its spatial position. Indeed, the velocity becomes ill-defined when particles' inertia causes the particles to detach from the flow and move ballistically destroying the fractal structure. This way of dealing with the problem of structure formation was introduced in \cite{FFS1} and developed for the problem with gravity in \cite{FHa}. 

We demonstrate that the condition that the velocity field is definable \cite{FHa} and thus the fractal is formed is $\min[St, Fr]\ll 1$. This transforms to the known condition $St\ll 1$ at zero gravity. Thus gravity allows to have situations where the velocity field is definable even at $St\gg 1$, where the particle's velocity relaxation time is comparable to the characteristic time-scale of turbulent eddies in the inertial range. Gravity damps particles' detachment from the flow \cite{FP1}. 

In the case of $St\gtrsim 1$ the particles' velocity field is not close to the velocity field of the ambient fluid and it cannot be expressed in terms of the latter in the closed form. We show that one can prove the existence of the velocity field, even not being able to write down an explicit expression for it. While it could seem that the mere existence of the field is not sufficient to reach definite conclusions, we show that this is not the case. We observe that if the velocity field is definable, then it must be weakly compressible, cf. \cite{FHa}.  This observation allows us to use the recent results on the universal properties of the stationary distributions of particles in the weakly compressible velocity field \cite{Fouxon2}. Those demonstrate that one can describe completely the statistics of particles' density in terms of just one dimensionless number $D_{KY}$ characterizing the statistics of turbulence. The analysis involves no modelling and gives testable predictions. We confirm the predictions by numerical simulations.

Determination of the range of $St$ and $Fr$ for which the particles form random fractal in space is basic question on the inertial particles' behaviour in turbulence. Remind that when gravity is negligible, $Fr\gg 1$, the fractal holds at $St<St_{cr}$ where $St_{cr}\sim 1$, while at larger $St$ the particles fill the whole space, albeit inhomogeneously \cite{Bec}. Qualitatively, if particles stick to the flow too much, $St\ll 1$, the flow mixes them efficiently, while if they are too heavy, $St\gg 1$, they move almost ballistically ignoring the action of vortices on them and tending to spread uniformly. We observe that the range of $St\ll 1$ for which one can define the particles' velocity field is asymptotically the same as the one for which the particles form fractal in space. This is because if the velocity field is definable, then it is compressible implying the existence of the fractal. Using similar considerations at $Fr>0$, our result suggests that the fractal exists at $\min[St, Fr]\lesssim 1$ and does not exist at $\min[St, Fr]\gg 1$. We provide the spectrum of fractal dimensions at $\min[St, Fr]\ll 1$, consider the limits of weak and strong gravity and give a number of experimentally testable predictions some of which we confirm numerically. 

\section{Basic equations and properties}
\label{bas}

We consider small spherical particle with radius $a$ and material density $\rho_p$ suspended in the fluid with density $\rho_f\ll \rho_p$ and kinematic viscosity $\nu$. The fluid flow $\bm u(t, \bm x)$ is assumed to be incompressible.  We assume that the particle's position $\bm x(t)$ and velocity $\bm v(t)$ obey the single-particle approximation where one can neglect the interaction between the particles and their back reaction on the flow. Further it is taken that the particle is sufficiently small so the fluid's force on the particle can be described by linear Stokes' drag. The consideration of the Reynolds number of the flow perturbation due to the drift of the water droplet in the warm cloud performed below demonstrates that the linear drag law holds well at droplet sizes smaller than $50\mu m$. 

The Newton law governing the evolution of the particle position $\bm x(t)$ and the particle velocity $\bm v(t)$ takes the form
\begin{equation}\label{Newton}
\frac{d \bm v(t)}{dt} =	 \frac{\bm u[t, \bm x(t)] - \bm v(t)}{\tau} + \bm g,
\end{equation}
where $\tau$ is the Stokes relaxation time of the particle's velocity, $\tau=2 \rho_p a^2/9 \nu \rho_f$, see \cite{MaxeyRiley, Maxey}.

After transients, the particle velocity is given by
\begin{equation}\label{vel_point}
\bm v(t) =
\bm v_g +
\frac{1}{\tau} \int_{-\infty}^{t} \bm u[t', \bm x(t')] exp \left( \frac{t'-t}{\tau} \right) dt',
\end{equation}
where we introduced the settling velocity in the absence of turbulence, $\bm v_g = \tau \bm g$. Using this equation one can write the drift velocity $\bm v_{d}(t)=\bm v(t)-\bm u[t, \bm x(t)]$ as 
\begin{equation}\label{drift}
\bm v_{d}\!=\!
\bm v_g\!+\!
 \int_{-\infty}^{t}\frac{dt'}{\tau}\left(\bm u[t', \bm x(t')]\!-\!\bm u[t, \bm x(t)]\right)\exp \left( \frac{t'-t}{\tau} \right) ,\nonumber
\end{equation}
where the last term describes the drift due to the inability of the particle to follow turbulent eddies with time-scale smaller than reaction time $\tau$. 

The smallest characteristic time-scale of turbulence is $\tau_\eta$ which is the characteristic time-scale of the turbulent fluctuations at the smallest scale of spatial variations of turbulence, the Kolmogorov scale $\eta$ \cite{Frisch}. It obeys $\tau_{\eta}= \sqrt{\nu/\epsilon}$ where $\nu$ is the kinematic viscosity and $\epsilon$ is the viscous energy dissipation per unit volume. The characteristic value of the turbulent velocity gradients is $\sqrt{\epsilon/\nu}=\tau_{\eta}^{-1}$. 

If the Stokes number defined by the ratio of $\tau$ and $t_{\eta}$ obeys $St=\tau/\tau_{\eta}\ll 1$ then there are no eddies with time-scale smaller or of the order of $\tau$. Thus if gravity can be disregarded then the particle would follow the flow closely with $\bm v_d\approx -\tau[\partial_t\bm u+(\bm u\cdot\nabla)\bm u]$, see \cite{Maxey}, so that $v_d\sim \eta  St/\tau_{\eta}$. When gravity is relevant one can write $v_{d}\sim \max [v_g, \eta  St/\tau_{\eta}]$.

In contrast if $St\gtrsim 1$ then there are turbulent eddies with time-scale smaller or of the order of the reaction time $\tau$ producing the drift velocity which is roughly the characteristic velocity of the eddies with time-scale $\tau$, see \cite{FH} for detailed discussion. Within the Kolmogorov theory where all properties in the inertial range are determined by $\epsilon$ the characteristic velocity of the eddies with time-scale $\tau$ is $\sqrt{\epsilon \tau}\sim \eta \sqrt{St}/\tau_{\eta} $. It follows that the drift is dominated by the largest of the contributions $v_g$ and $\eta \sqrt{St}/\tau_{\eta}$ so one has $v_{d}\sim \max [v_g,\eta \sqrt{St}/\tau_{\eta} ]$. Similarly one can consider the case of $\tau$ which is much larger than the integral time-scale of turbulence which is beyond our scope here. 

It follows that the minimal timescale of variations of the turbulent velocity $\bm u[t, \bm x(t)]$ in the particle reference frame is of order $\tau_c = \min[\tau_\eta, \tau_{d}]$. Here $\tau_{d}=\eta/v_d$ is the time during which the particle drifts through that smallest scale of spatial variations of turbulence. 
If the drift is dominated by gravity, $v_g\gtrsim  \eta  St/\tau_{\eta}$ at $St\lesssim 1$ or  $v_g\gtrsim  \eta  \sqrt{St}/\tau_{\eta}$ at $St\gg 1$, then $\tau_d=\tau_g$ where $\tau_g = \eta / v_g$. 

The time-scale $\tau_g$ defines the remaining dimensionless parameter of the problem, the Froude number, $Fr \equiv \tau_g \tau / \tau_\eta^2$. Note that this parameter is independent of the parameters of the particles: $Fr=\eta/[g\tau_\eta^2]$ describes the ratio of $\eta$ to the distance passed by the particle accelerating at $g$ within the viscous time-scale $t_{\eta}$ (this definition of the Froude number is traditional in the area, see \cite{review} and references therein). One can write $Fr=\epsilon^{3/4}/[g\nu^{1/4}]$. 

The Froude number is independent of the parameters of the particles so it can be used to talk of various situations of turbulence irrespective of the particles. We consider the case of turbulence in warm clouds using the air viscosity $\nu=1.5*10^{-1} cm^2/s$. One finds $Fr\sim 2\epsilon^{3/4}*10^{-3}$ where $\epsilon$ is measured in $cm^2/s^3$. Using as the upper bound $\epsilon=10 cm^2/s^3$ for stratocumulus clouds and  $\epsilon=10^2 cm^2/s^3$ for cumulus ones, see \cite{review} and references therein, one finds $Fr \sim 0.01$ for stratocumulus clouds and $Fr \sim 0.06$  for cumulus clouds. Thus the Froude number typical for warm clouds is very small signifying large role of gravity (the other dimensionless parameter $St$ grows quadratically with the particle radius. For droplets with size $15 \mu m$ one has $St\sim 0.07$). We will see that the smallness of $Fr$ guarantees that multi-streaming of particles has low probability so that there is a well-defined flow of particles. In contrast, using for clouds $\epsilon=10^3 cm^2/s^3$ considered in \cite{FP1} one finds $Fr\sim 0.3$ which is not small so it is likely that the multi-streaming is rather typical and there is no well-defined particles' flow. In fact it was found in \cite{FP1} that the sling effect is relevant so that the spatial motion of the particles cannot be described by flow. It seems that $\epsilon=10^2 cm^2/s^3$ is a rough upper limit for the possibility to introduce the flow of particles. In the discussion below we consider clouds with $\epsilon\lesssim 10^2 cm^2/s^3$ which seems to fit the typical situations see \cite{review} and references therein. 

Similar considerations can be performed in water where using $\nu=10^{-2} cm^2/s$ one has $Fr\sim 3\epsilon^{3/4}*10^{-4}$ where $\epsilon$ is measured in $cm^2/s^3$. In this case significantly smaller $\epsilon$ than those considered above hold in typical aquatic environments. The consideration of the corresponding physical cases is beyond the scope of this work. 

Using $St$ and $Fr$ the condition that the drift is dominated by gravity, $\bm v_{d}\approx \bm v_g$ can be written as $Fr\ll 1$ at $St\lesssim 1$ and 
$St^{-1/2}Fr\ll 1$ at $St\gg 1$ where we observed that $Fr=\eta  St/v_g \tau_{\eta}$. Thus in cases where $Fr\ll 1$ the drift is dominated by gravity at arbitrary Stokes number. In particular we conclude that $\bm v_{d}\approx \bm v_g$ for water droplets in warm clouds. 

We now consider the applicability of the linear friction law to the motion of water droplets in clouds. The Reynolds number based on the drift velocity $\bm v_g$ is given by $v_g a/\nu= 2 g\rho_p a^3/[9 \nu^2 \rho_f]$. Using $\rho_p/\rho_f=784$, $\nu=1.5*10^{-1} cm^2/s$ we find $v_g a/\nu=(a/50\mu m)^3$.
Since the size dependence is quite strong then the use of the linear friction law for sizes smaller than $50 \mu m$ seems valid. 

The remaining dimensionless parameter, the Reynolds number $Re$, determines the ratio of the gravitational settling velocity $v_g$ to the large-scale velocity of turbulence. Using the Kolmogorov theory for estimates we have $u_L\sim (\epsilon L)^{1/3}=(\epsilon \eta)^{1/3}(L/\eta)^{1/3}=\eta Re^{1/4}/\tau_{\eta}$ where we use that the Reynolds number at the viscous scale $\epsilon^{1/3}\eta^{4/3}/\nu$ is of order one so that $Re =\epsilon^{1/3}L^{4/3}/\nu=(L/\eta)^{4/3}$. We observed that the characteristic velocity $\sqrt{\epsilon \tau_{\eta}}$ of turbulence at the viscous time-scale is given by $\eta/\tau_{\eta}$.
It follows that $v_g/u_L=Re^{-1/4}\tau_{\eta}/\tau_g=Re^{-1/4}St Fr^{-1}$. For clouds $Re^{1/4}$ ranges from $\sim 70$ in stratocumulus clouds to $\sim 200$ in cumulus ones where we estimated $Re$ as the square root of the Taylor-scale Reynolds number, see \cite{review} and references therein. Using the previous estimates for $Fr$ we find that $Re^{-1/4}Fr^{-1}\sim 1.4$ for stratocumulus clouds and $Re^{-1/4}Fr^{-1}\sim 0.1$ for cumulus ones. 
Thus we have $v_g/u_L\sim St$ for stratocumulus clouds and $v_g/u_L\sim 0.1 St$ for cumulus clouds. 

We observe that the situation of $v_g/u_L\gtrsim 1$ is possible. In this case the gravitational settling of the particles at the speed $\bm v_g$ constitutes significant part of their spatial motion. In particular, the situation $v_g/u_L\gg 1$ is quite possible where turbulence provides small corrections to the settling. Nevertheless we will see that the preferential concentration of the particles occurs predominantly at scales smaller than $\eta$. One has to distinguish the instantaneous pattern of spatial motion where the particles can concentrate in particular regions due to peculiarities of the large-scale motion from the persistent preferential concentration in particular flow regions. In particular, the theory described below is largely independent of whether $v_g/u_L$ is smaller or larger than $1$ though the overall pattern of the particles' motion is strongly different in the two cases. 

\section{Flow of particles can exist without explicit formula for it}

The approach of this work to the solution of the particle's equations of motion (\ref{Newton}) is via the introduction of the spatial flow. It is instructive to consider first the particular case of $\tau \ll \min[\tau_\eta, \tau_g]$. In this case $\tau$ is much smaller than the characteristic time-scale of variations of $\bm u[t,\bm x(t)]$ in the integrand of Eq. (\ref{vel_point}) since the latter is given by the minimum of $\min[\tau_\eta, \tau_g]$ and the inertial drift time-scale $\tau_{\eta}/St$ (we use that $\tau \ll \min[t_\eta, \tau_g]$ implies $St\ll 1$ so inertial drift velocity is of order of $St v_{\eta}\sim St \eta/\tau_{\eta}$). Thus one can write the expansion of the particle's velocity in powers of $\tau$ which can be obtained by using Taylor expansion of $\bm u[t,\bm x(t)]$ in the integrand of Eq. (\ref{vel_point}). To order $\tau$ one finds \cite{Maxey}
\begin{equation}\label{eff_vel}
\bm v(t) \approx
\bm v_g + \bm u
- \tau \left[\partial_t \bm u + (\bm u \cdot \bm\nabla) \bm u + (\bm v_g \cdot \bm\nabla) \bm u \right],
\end{equation}
where $\bm u$ is evaluated at $[t, \bm x(t)]$.
The particle's velocity is uniquely determined by its position in space, $\bm v(t) = \bm v[t, \bm x(t)]$, where $\bm v(t, \bm x)$ is prescribed velocity field $\bm  v=\bm v_g + \bm u - \tau \left[\partial_t \bm u + (\bm u \cdot \bm\nabla) \bm u + (\bm v_g \cdot \bm\nabla) \bm u \right]$. The term in brackets in Eq.~(\ref{eff_vel}) presents small correction to $\bm v_g + \bm u$ because $St\ll 1$. This correction is the small-scale one because both the field of accelerations $\partial_t \bm u + (\bm u \cdot \bm\nabla) \bm u$ and gradients $\nabla \bm u$ are determined by the small-scale turbulence with characteristic scale $\eta$. Despite that this correction to absolute velocity $\bm v(t, \bm x)$ is small, its correction to $\nabla\cdot \bm v$ is small in no way - $\nabla\cdot \left[\bm v_g + \bm u\right]$ is identically zero while $\nabla\bm v=-\tau\phi$ where $\phi=(\nabla_ku_i)(\nabla_iu_k)$. Thus all changes in the particles' density (besides transport) are solely due to the small, small-scale correction $\tau \left[\partial_t \bm u + (\bm u \cdot \bm\nabla) \bm u + (\bm v_g \cdot \bm\nabla) \bm u \right]$. That creates highly non-trivial density structure at the viscous scale having non-perturbative effect on the particles' density \cite{Fouxon2}. 

The latter structure is to be distinguished clearly from the spatial pattern of particles' motion due to the turbulent transport that, of course, can be highly non-trivial too. The distinguishing occurs when one covers the space uniformly with particles. The uniform density would not be changed by the turbulent transport but it is changed to fractal structure by the $\tau \left[\partial_t \bm u + (\bm u \cdot \bm\nabla) \bm u + (\bm v_g \cdot \bm\nabla) \bm u \right]$ term, see the next Section. When working with (necessarily) finite number of particles one has to separate patterns of clustering from the ones of transport.

The possibility to introduce in a range of parameters the flow of particles $\bm v(t, \bm x)$ that describes the particles' motion in space by 
\begin{equation}\label{basic}
\dot {\bm x} = \bm v[t, \bm x(t)],
\end{equation}
is a significant simplification \cite{Maxey, FFS1, Fouxon2}. This description says that the particle's velocity is determined uniquely by its spatial position so there is no multi-streaming at the same point. Clearly due to arbitrariness in prescribing the initial position and velocity of the particles this description holds only after transients, cf. Eq.~(\ref{vel_point}) and the derivation of this Section. 

It is usually impossible to introduce the flow of particles fulfilling Eq.~(\ref{basic}) since Eq.~(\ref{Newton}) is the flow in the phase space whose projection on the real space is typically multi-valued with possibility of different velocities of the particle at the same spatial point. But if the description (\ref{basic}) does hold then a lot of information on the particles' behavior becomes available. Remarkably this information is universal 
that is a whole range of properties of the spatial distribution of particles obeying Eq.~(\ref{basic}) is independent of the explicit form of $\bm v(t, \bm x)$. 

Our approach in this work is to determine the range of parameters where the description \eqref{basic} with certain effective velocity $\bm v(t, \bm x)$ holds. The conventional study of this question, briefly reviewed in the next Section, relies on the explicit solution (\ref{eff_vel}) where the description by flow is proved by explicit construction, cf. \cite{FHa}. This however misses the possibility that Eq.~\eqref{basic} can hold in the absence of explicit expression for $\bm v(t, \bm x)$. We will demonstrate that indeed there are cases where the flow exists but the explicit formula for it does not. 

\section{Universal particles distribution in weakly compressible flow}\label{theory}

It was demonstrated in the recent work \cite{Fouxon1} that the spatial distribution of particles obeying Eq.~(\ref{basic}) where $\bm v(t, \bm x)$ is weakly compressible smooth flow has universal properties independent of details on the flow, see also \cite{Fouxon2}. The involved assumptions are 
finite correlation time of the gradients of $\bm v(t, \bm x)$ and the spatial uniformity of the gradients' statistics. The former assumption usually holds for turbulence while the latter holds far from the boundaries. We stress that the discussed phenomena occur at the Kolmogorov scale $\eta\ll L$ so the assumption of the spatially uniform statistics is typically obeyed well. 

In the steady state the particles distribute over fractal set in space in contrast to the uniform distribution holding for incompressible mixing flow. This set can be characterized uniquely by one number - the Kaplan-Yorke codimension $D_{KY}$. 
Since the flow is weakly compressible then the usual definition \cite{KY} of $D_{KY}$ reduces to $D_{KY} = \sum \lambda_i / \lambda_3$, see details in Sec. \ref{fractaldimensions}.
Here $\sum \lambda_i < 0$ is the sum of the Lyapunov exponents \cite{reviewp} of the flow $\bm v$, which determines the average logarithmic growth (or rather decrease) rate of infinitesimal volumes, see Sec. \ref{sumsection}. The exponent $|\lambda_3|$ in the weakly compressible case is the backward-in-time logarithmic rate of separation of infinitesimally close particles, see Sec. \ref{third}. The weak compressibility implies $D_{KY}\ll 1$.

The trajectories of the particles in space asymptote a multi-fractal set at large times. Correspondingly, the density of the inertial particles, supported on the set with zero spatial volume, becomes singular at large times. The statistics of the resulting singular steady state density $n_s$ are lognormal. They are described by (we use normalization with $\langle n_s\rangle=1$)
\begin{equation}\label{density_corr}
\left\langle n_{s}(\bm x_1) n_{s}(\bm x_2) \ldots n_{s}(\bm x_N) \right\rangle \! =
\exp \!\left[ \sum_{i>j} g(\bm x_i \! - \! \bm x_j) \right]\!.
\end{equation}
Thus all the correlation functions can be found using the pair-correlation function $\left\langle n_{s}(0)n_{s}(\bm x) \right\rangle = \exp \left[ g(\bm x) \right]$. The function $g(\bm x)$ has a universal logarithmic behavior at small scales \cite{Fouxon1} which gives
\begin{equation}\label{corr_ident}
\left\langle n_s(0) n_s(\bm x) \right\rangle = \left( \frac{\eta}{x} \right)^{2D_{KY}},
\ \
x \ll \eta.
\end{equation}
We observe that the correlation dimension - the exponent of the power-law in the formula above - is twice the Kaplan-Yorke dimension. In fact the latter dimension determines the complete spectrum of fractal dimensions $D(\alpha)$ defined by \cite{HP, BGH}
\begin{equation}
D(\alpha) \equiv \lim_{l \to 0} \ln \frac{\left\langle m_l^{\alpha-1} n_{s} \right\rangle}{(\alpha - 1) \ln l} \ ,
\end{equation}
where $m_l$ is the mass in the small ball, closely related to the coarse-grained density $n_l$,
\begin{equation}
m_l(\bm x) \equiv \int_{|\bm x'-\bm x|<l} n_s(\bm x') d\bm x'
, \ \
n_l(\bm x) \equiv \frac{3m_l(\bm x)}{4\pi l^3} \ . 
\end{equation}
One has \cite{Fouxon1}
\begin{equation}
D(\alpha) = 3 - D_{KY} \alpha,\label{spectrumdimensions}
\end{equation}
while $n_l(\bm x)$ has lognormal statistics. We observe that the fractal dimension $D(0)$ coincides with the dimension of space $3$ (this does not contradict multifractality) and the information dimension $D(1)$ is equal to the Kaplan-Yorke dimension. Since $D_{KY} \ll 1$ then $\left\langle n_s(0) n_s(\bm x) \right\rangle\approx \left\langle n_s(0) \right\rangle \left\langle n_s(\bm x) \right\rangle =1$ unless $x$ is significantly smaller than $\eta$, see Eq.~(\ref{corr_ident}). Thus the density fluctuates significantly only at scales much smaller than $\eta$ and equation \eqref{corr_ident} completely determines the statistics in the region where the fluctuations are significant. 

Finally we indicate useful way to write $D_{KY}$ \cite{Fouxon1, Fouxon2}. This uses the Green-Kubo type formula for the sum of Lyapunov exponents derived in \cite{FF}
\begin{eqnarray}&&\!\!\!\!\!\!\!
\sum \lambda_i=-\int_0^{\infty} \langle \nabla\cdot\bm v(0, \bm x_0) \nabla\cdot\bm v[t, \bm x(t, \bm x_0)]\rangle dt,\label{sumlambda}
\end{eqnarray} 
where the average is the spatial average over $\bm x_0$ and $\bm x(t, \bm x_0)$ is the trajectory of the particle issuing from $\bm x_0$,
\begin{eqnarray}&&\!\!\!\!\!\!\!
\partial_t \bm x(t, \bm x_0)=\bm v\left[t, \bm x(t, \bm x_0)\right],\ \ \bm x(t=0, \bm x_0)=\bm x_0.
\end{eqnarray} 
It follows that 
\begin{eqnarray}&&\!\!\!\!\!\!\!\!\!\!\label{ky}
D_{KY}=\frac{1}{|\lambda_3|}\int_0^{\infty} \langle \nabla\cdot\bm v(0, \bm x_0) \nabla\cdot\bm v[t, \bm x(t, \bm x_0)]\rangle dt.
\end{eqnarray}
The universality signifies that different weakly compressible flows with finite correlations in time produce the same statistics of the particles' distribution in space. The only difference between different flows is in the value of $D_{KY}$. The origin of this remarkable universality is in the trivial universality of incompressible mixing flows - whatever the statistics of that flow is, in the steady state the distribution of particles is uniform in space. 

First application of the above theory was to the study of the distribution of weakly inertial particles with $St\ll 1$ in the Navier-Stokes turbulence when gravity can be neglected \cite{Fouxon1}. This is the case of Eq.~(\ref{eff_vel}) with $\bm v_g=0$. Since compressibility comes from the small correction to velocity then we deal with Eq.~(\ref{basic}) where $\bm v$ is weakly compressible and has finite correlations in time (because turbulence does). Due to universality the only thing that remains to discuss is the value of $D_{KY}$. Using Eq.~(\ref{ky}) and $\nabla\cdot \bm v=-\tau\phi$ one finds 
\begin{equation}
D_{KY} = \frac{\tau^2}{2|\lambda_3|} \int_{-\infty}^{\infty} \left\langle \phi[0, 0] \phi[t, \bm x_u(t, 0)] \right\rangle dt,\label{kydim01}
\end{equation}
where to leading order $\lambda_3$ is the third Lyapunov exponent of turbulence and $\bm x_u(t, 0)$ are the usual Lagrangian trajectories of the fluid particles,
\begin{equation}\label{Lagrangian}
\partial_t \bm x_u(t, \bm x)=\bm u[t, \bm x_u(t, \bm x_0)],\ \ \bm x_u(t=0, \bm x_0)=\bm x_0.
\end{equation}
We noted that incompressibility of turbulence implies that $\left\langle \phi[0, 0] \phi[t, \bm x_u(t, 0)] \right\rangle$ is even function of $t$ so that one can integrate over the domain $(-\infty, \infty)$ rather than $(0, \infty)$ in Eq.~(\ref{ky}). It is readily seen from Eq.~(\ref{kydim01}) that $D_{KY} \propto St^2 \ll 1$, cf. \cite{FFS1}.
Note that $\left\langle \phi[0, 0] \phi[t, \bm x_u(t, 0)] \right\rangle$ is the usual definition of different time correlation function and $-\phi$ is equal to the Laplacian of pressure as can be seen by taking the divergence of the Navier-Stokes equations. 

Though the consideration was performed \cite{Fouxon1} neglecting gravity it is straightforward to include the gravity in the case  $\tau \ll \min[\tau_\eta, \tau_g]$ described by Eq.~(\ref{eff_vel}), cf. \cite{FHa}. The corresponding change in $D_{KY}$ is 
\begin{equation}
D_{KY} = \frac{\tau^2}{2|\lambda_3^w|} \int \left\langle \phi[0, 0] \phi[t, \bm x_w(t, 0)] \right\rangle dt,\label{kydim1}
\end{equation}
where $\lambda_3^w$ is the third Lyapunov exponent of the incompressible flow $\bm w\equiv \bm u+\bm v_g$ and the trajectories $\bm x_w(t, \bm x)$ are defined by ($\bm x_w(0, \bm x_0)=\bm x_0$)
\begin{eqnarray}
&&\!\!\!\!\!\!\!\!\!\!\!\!\!\!\!\! \partial_t \bm x_w(t, \bm x_0)=\bm w\left[t,  \bm x_w(t, \bm _0x)\right]=
\bm u\left[t,  \bm x_w(t, \bm x_0)\right]+\bm v_g.  \label{Lagr}
\end{eqnarray}
When $\bm v_g$ can be neglected Eq.~(\ref{ky}) is reproduced. Since Eq.~(\ref{kydim1}) seems to be studied previously only in the case $\bm v_g=0$ then we will consider it in detail below. 

Further use of the universality came in the late work \cite{Nature2013} where the same theoretical predictions were used to describe the distribution of living phytoplankton cells in the ocean. This is though the microscopic equations governing the inertial particles and phytoplankton are rather different. In particular range of parameters the cells' motion in space is determined by weakly compressible flow with finite correlations so that the described predictions hold with the corresponding $D_{KY}$. 

The present work provides further use of universality that goes one step beyond the previous uses in that it works with weakly compressible flow in the range of parameters where no explicit formula for $\bm v$ similar to Eq.~(\ref{eff_vel}) holds. Rather we prove indirectly that the flow of particles can be defined and it is weakly compressible without knowing the expression of it, cf. \cite{FHa}. It is to be stressed that the difference is not the technical one of deriving the formula - rather the particles' flow dependence on $\bm u(t, \bm x)$ is no longer local in time that is instantaneous turbulent flow $\bm u(t, \bm x)$ does not determine the instantaneous particles' flow $\bm v(t, \bm x)$.

Once the existence of weakly compressible particles' flow is proved the problem is reduced to one unknown parameter - $D_{KY}$. We manage to determine $D_{KY}$ reducing the whole spectrum of the Lyapunov exponents to one number - the integral of the energy spectrum times the wavenumber. 

To understand the reasons behind the possibility to introduce the flow of particles consider first the case when gravity can be disregarded (the limit $Fr=\infty$). Then the flow cannot be introduced if $St\gtrsim 1$ but can be introduced if $St\ll 1$. The condition $St\ll 1$ or $\tau\ll \tau_{\eta}$ guarantees that the typical turbulent vortex (understood here as particular configuration of the flow gradient) varies over time-scale much larger than the reaction time. There are however strong vortices in the turbulent flow whose characteristic time-scale is of order $\tau$. These vortices instead of rotating the particle that slightly deviates from the local motion of the fluid would forcefully throw the particle outside like in the sling.
This produces the so-called sling effect introduced in \cite{FFS1} and recently confirmed experimentally in \cite{Bewley}. 

The motion of particles thrown out by strong vortices is determined by inertia so it is essentially ballistic. The particles move ballistically colliding with other particles on their way. We stress that these collisions are not due to the finite size of the particles but rather these are collisions of point particles due to intersections of different streams of particles in the same spatial region. In particular, the corresponding rate of collisions decays slower in the limit of zero particle size. It was demonstrated that though the sling effect is rare event at $St\ll 1$ it still has to be included in the overall rate of collisions because slings create optimal conditions for collisions \cite{FFS1}. When $St\sim 1$ the sling effect becomes typical so that the flow of particles does not exist if $St\gtrsim 1$. There is no well-defined flow of particles $\bm v(t, \bm x)$ and the motion of particles in space is not described by Eq.~(\ref{basic}). 

When gravity is present it becomes possible to consider the possibility that the flow of particles can be introduced though $St\gtrsim 1$. The reason is that for sling effect to hold, given turbulent vortex needs to act coherently on the inertial particle during the time-scale of order $t_{\eta}$ so that the end result is the ballistic throwing of the particle out of the vortex. Gravitational fall of the particle through the vortex during the time-scale $\tau_g$ can shorten the time of the action of the vortex on the particle so that no sling effect occurs. 
In the next Section we demonstrate that due to gravity the flow of particles can in fact be introduced at $St\gtrsim 1$ provided that $Fr\ll 1$. 

\section{Flow of particles can be introduced when $St\gtrsim 1$ if $Fr\ll 1$}

In this Section we determine the range of $St$ and $Fr$ when the flow of particles can be introduced. We observe that the study has to follow a different route from the one taken at $St\ll 1$. This is because it is not possible to produce formula like Eq.~(\ref{eff_vel}) that provides the velocity of particle $\bm v(t)$ in terms of the turbulent flow at the same moment of time $t$. Rather at $St\gtrsim 1$ the time of variations of $\bm u[t', \bm x(t')]$ in the integral in Eq.~(\ref{vel_point}) is smaller or equal to $\tau$ (indeed that time is at least $t_{\eta}$ which is of order of $\tau$ or smaller at $St\gtrsim 1$). Thus at $St\gtrsim 1$ the particle's velocity $\bm v(t)$ is determined by the turbulent flow not only at time $t$ but also at the previous moments of time so that the dependence of $\bm v(t)$ on $\bm u(t, \bm x)$ is non-local in time. We cannot introduce the flow of particles by constructing it via a formula similar to Eq.~(\ref{eff_vel}) when $St\gtrsim 1$.

In the case of $St\gtrsim 1$ we demonstrate the existence of the flow of particles not by constructing it explicitly but rather by proving its self-consistency when $Fr\ll 1$. We assume that the solution to equation \eqref{Newton} after transients obeys \eqref{basic} with certain $\bm v(t, \bm x)$. If so then the differentiation of equation \eqref{basic} gives 
\begin{eqnarray}
&& \frac{d\bm v}{dt}=\partial_t\bm v+(\bm v\cdot\nabla)\bm v. 
\end{eqnarray}
Comparing this equation with Eq.~\eqref{Newton} we conclude that $\bm v(t, \bm x)$ must obey the PDE
\begin{equation}\label{eff_vel_cond}
\partial_t \bm v + (\bm v \cdot \bm\nabla) \bm v =
\frac{\bm u - \bm v}{\tau} + \bm g,
\end{equation}
cf. \cite{FFS1,FHa}. Then the condition of self-consistency of the assumption \eqref{basic} is that the velocity field $\bm v(t, \bm x)$ evolving according to equation (\ref{eff_vel_cond}) remains well-defined at all times. Indeed, consider the initial conditions where particles are distributed in space so that their initial velocity obeys $\bm v(t=0)=\bm v[t=0, \bm x(t=0)]$ where $\bm v(t=0, \bm x)$ is a smooth field. The general case of evolution prescribed by \eqref{Newton} is that at a certain moment of time $t_*$ for the first time two particles would come to the same spatial point having different velocities (when the projection on the position space of the unique trajectories in the phase space stops to be single-valued) signifying the breakdown of the effective description \eqref{basic}. This breakdown will be signalled to us by the divergence of the velocity gradients at $t = t_*$ due to the possibility to have finite difference of $\bm v(t, \bm x)$ at points separated by zero difference of the coordinates. We conclude that the self-consistency of the flow description \eqref{basic} of the solutions to equation \eqref{Newton} demands that there is no finite time blow up of the velocity gradients of $\bm v$ obeying equation \eqref{eff_vel_cond}. 

To study the blow up we observe that the gradients $\sigma_{ij}(t, \bm x) \equiv \nabla_j v_i(t, \bm x)$ satisfy the matrix partial differential equations (PDE) \cite{FHa,FFS1}
\begin{equation}\label{vel_gradients1}
\partial_t \sigma + (\bm v \cdot \bm\nabla) \sigma + \sigma^2 =
\frac{s - \sigma}{\tau},
\end{equation}
where we define $s_{ij}(t, \bm x) \equiv \nabla_j u_i(t, \bm x)$.
In the particle's frame the gradients $\sigma(t)=\nabla_j v_i[t, \bm x(t)]$ obey the ordinary differential equations (ODE)
\begin{equation}\label{vel_gradients2}
\frac{d \sigma}{dt} + \sigma^2 =-
\frac{\sigma}{\tau}+\frac{s}{\tau}.
\end{equation}
We observe that the gradients $\sigma$ are produced by the gradients $s$ of turbulence in the particle's frame which are finite. If those gradients produce $\sigma\ll 1/\tau$ then the non-linear $\sigma^2$ term is much smaller than the damping term $-\sigma/\tau$ so that the gradients $\sigma$ obey after transients
\begin{equation}\label{linear}
\sigma \approx \sigma_l,
\ \
\sigma_l \equiv \frac{1}{\tau} \int_{-\infty}^{t} s(t') \exp \left( \frac{t'-t}{\tau} \right) dt',
\end{equation}
where the subscript $l$ stands for linear. Clearly in this case $\sigma$ are finite so that the flow description \eqref{basic} is self-consistent. On the contrary if $s$ produces $\sigma_l\gtrsim 1/\tau$ then the non-linear $\sigma^2$ term in Eq.~(\ref{vel_gradients2}) starts to dominate the dynamics producing a finite-time blow up of $\sigma(t)$ (because solutions to $\dot{\sigma}+\sigma^2=0$ blow up in finite time $t_c$ as $(t-t_c)^{-1}$). We conclude that the condition of self-consistency of the effective description \eqref{basic} is $\left\langle \sigma_l^2 \right\rangle \tau^2\ll 1$. 

Observing from Eq.~(\ref{linear}) that $\sigma_l\sim \int_{t-\tau}^t s(t')dt'/\tau$ we obtain $\left\langle \sigma_l^2 \right\rangle \sim \tau^{-1} \int_{0}^{\tau} \left\langle s(0)s(t) \right\rangle dt \sim \min[\tau_c, \tau] (\tau_\eta^2 \tau)^{-1}$ where we introduced the correlation time $\tau_c$ of $s(t)$ and noticed that for turbulence $\left\langle s^2 \right\rangle \sim \tau_\eta^{-2}$. Since $\tau_c=\min[\tau_{\eta}, \tau_{d}]$ then $\left\langle \sigma_l^2 \right\rangle \sim \min[\tau, \tau_\eta, \tau_{d}](\tau_\eta^2 \tau)^{-1}$. We conclude that the effective velocity field description \eqref{basic} holds provided
\begin{eqnarray}&&\tau \min[\tau, \tau_\eta, \tau_{d}]\ll \tau_\eta^2.\label{condition}
\end{eqnarray}
The discussion in this Section involved no limitations on $St$, $Fr$ so far. We now concentrate on the case $St\gtrsim 1$ considering different ranges of parameters. 

If $St^{-1/2}Fr\gtrsim 1$ then the drift velocity is due to turbulence and not gravity, $v_{drift}\sim \sqrt{\epsilon \tau}$, see Section \ref{bas}. In this case $\tau_{d}\sim \eta/ \sqrt{\epsilon \tau}\sim \tau_{\eta}St^{-1/2}$. This time-scale is smaller than $\tau_{\eta}$ so that it equals $\tau_c$. We find $\tau_c=\tau_{\eta}St^{-1/2}\lesssim \tau$. 
The condition (\ref{condition}) then gives $St^{1/2}\ll 1$ that contradicts $St\gtrsim 1$. Thus in this case the assumption \eqref{basic} is inconsistent so the flow of particles cannot be introduced when $St\gtrsim 1$ and $St^{-1/2}Fr\gtrsim 1$. 

On the contrary, consider $St\gtrsim 1$ and $St^{-1/2}Fr\ll 1$. Then the drift velocity is due to gravity $\tau_{d}\sim \tau_g$ where $\tau_g/\tau=  Fr St^{-2}\ll 1$ and $\tau_g/\tau_{\eta}= Fr St^{-1}\ll 1$. In this case the condition (\ref{condition}) gives $Fr\ll 1$. Combining the inequalities we conclude that the flow of particles can be introduced at $St\gtrsim 1$ if $Fr\ll 1$, cf. \cite{FHa}. 

We make a technical remark. Observe that at $St\gtrsim 1$ and $Fr\ll 1$ we have $\langle\sigma_l^2\rangle \tau^2 \sim Fr$ so that $\sigma_l\tau\sim Fr^{1/2}$. Thus one can wonder if the true condition that the probability of the blow up is small is $Fr^{1/2}\ll 1$ rather than $Fr\ll 1$. The difference is significant - for water droplets in clouds $Fr\ll 1$ but $Fr^{1/2}$ could be not small (say for $Fr=0.05$ in cumulus clouds one has $Fr^{1/2}=0.22$ which smallness could be destroyed by a numerical factor of order one). It will become clear from Gaussianity demonstrated in the next Section that $Fr\ll 1$ is the true criterion.  

\section{Flow of particles at $St\gtrsim 1$ and $Fr\ll 1$ is weakly compressible with short-correlated derivatives}

It follows from the study in previous Sections that the flow of particles $\bm v(t, \bm x)$ can be introduced when $St\gtrsim 1$ and $Fr\ll 1$ but there is no closed formula for it in terms of $\bm u(t, \bm x)$. It could seem that not much can be said then on the distribution of particles in space. In fact the contrary holds - we can describe the statistics of particles' distributions in space completely based on one phenomenological constant that characterizes the statistics of the transporting turbulent flow. Further, that number can be written with the help of the single-time correlation functions of turbulence that permit direct comparison with the experiment. The simplification happens due to the observation that whenever the flow of particles can be introduced it is going to be weakly compressible (so that the results of Section \ref{theory} can be used) and the observation of Gaussianity of $\sigma$. 

We first demonstrate that the flow $\bm v(t, \bm x)$ is weakly compressible \cite{FHa}, that is in the decomposition of $\bm v(t, \bm x)$ into solenoidal and potential components the latter will include smallness parameter. To demonstrate this we observe that by construction $\sigma\approx \sigma_l$ when 
$St\gtrsim 1$ and $Fr\ll 1$. However the incompressibility of turbulence implies $tr s=0$ so that $tr\sigma_l=0$, see Eq.~(\ref{linear}). It follows that the leading order approximation $\sigma\approx \sigma_l$ that completely neglects the non-linear $\sigma^2$ term in Eq. (\ref{vel_gradients2}) is degenerate - it produces $\bm v$ whose compressibility is identically zero. The next order approximation however renders $\bm v$ finite compressibility.  

To find the leading order expression for $tr \sigma$ we rewrite equation \eqref{vel_gradients2} in the integral form
\begin{eqnarray}&&\!\!\!\!\!\!\!\!\!\!\!\!\!\!\!\!
\sigma(t)= \int_{-\infty}^t \exp \left( \frac{t'-t}{\tau} \right) \left[ \frac{s(t')}{\tau} + \sigma^2(t') \right] dt'=
\sigma_l(t)\nonumber
\\&&\!\!\!\!\!\!\!\!\!\!\!\!\!\!\!\!
+ \int_{-\infty}^t \exp \left( \frac{t'-t}{\tau} \right) \sigma^2(t') dt'.
\end{eqnarray}
where $\sigma_l$ is defined in equation \eqref{linear}.
Solving it by iterations we find \cite{FFS}
\begin{eqnarray}&&\!\!\!\!\!\!\!\!
\sigma(t)= \sigma_l(t) + \int_{-\infty}^t \exp \left( \frac{t'-t}{\tau} \right) \sigma_l^2(t') dt'+\ldots,
\end{eqnarray}
where dots stand for higher order terms in $\sigma_l \tau\ll 1$. Thus
\begin{equation}
tr \sigma(t) \approx \int_{-\infty}^t \exp \left( \frac{t'-t}{\tau} \right) tr \sigma_l^2(t') dt'. \label{trace}
\end{equation}
Since the time of variations of $\sigma_l(t)$ is $\tau$ or larger, cf. equation \eqref{linear}, then the above equation gives $tr \sigma(t) \sim 
\tau tr \sigma_l^2(t)$ which is much smaller than $\sigma\approx\sigma_l$ by $\sigma_l \tau\ll 1$.
We obtain that if the flow of particles exists then it is weakly compressible so that its divergence is much smaller than the typical value of it gradients. It follows that in the decomposition of $\bm v$ into solenoidal and potential components, 
\begin{equation}\label{weak_compr_cond}
\bm v = \bm v_{\perp} +\bm v_{||},
\ \
\nabla \cdot \bm v_{\perp} = 0,\ \ \nabla \times \bm v_{||} = 0.
\end{equation}
the potential component is much smaller than the transversal one, $v_{||}\ll v_{\perp}$. Of course the origin of this smallness is in the incompressibility of turbulence, $tr s=0$, though to trace it is non-trivial at $St\gtrsim 1$ due to temporal non-locality of the dependence of $\bm v(t, \bm x)$ on $\bm u(t, \bm x)$. In the familiar particular case of equation \eqref{eff_vel} where the considerations above hold too these conclusions follow directly by taking derivatives. 

We observed previously that $\left\langle \sigma_l^2 \right\rangle \sim Fr/\tau^2$ when $St\gtrsim 1$ and $Fr\ll 1$. It follows that the typical value of $\sigma$ is $Fr^{1/2}/\tau$. Further, the typical time $\tau_g$ of variations of $s$ is much smaller than $\tau$. It follows from equation \eqref{linear} that roughly $\sigma(t)$ is $s(t)$ coarse-grained over the time-scale $\tau$ which is much larger than the correlation time $\tau_g$ of $s(t)$. Thus $\sigma(t)$ is Gaussian (indeed $\sigma(t)$ is roughly the sum of large number $\sim \tau/\tau_g$ of random independent variables. Rigorous proof can be obtained by observing that all cumulants of $\sigma$ depend on $\tau$ linearly. Using then the cumulant expansion theorem \cite{Ma} to find the characteristic functional of $\sigma(t)$ one proves the Gaussianity). It follows that the probability that $|\sigma|\geq 1/\tau$ is proportional to the Gaussian weight $\exp[-(2\tau^2\langle \sigma_l^2\rangle)^{-1}]$ confirming the criterion used in the previous Section that the probability that $|\sigma|\geq 1/\tau$ is rare event if $\tau^2\langle \sigma_l^2\rangle\ll 1$. 
Further the product of the typical value of $\sigma(t)$ and its correlation time $\tau$ is given by $Fr^{1/2}\ll 1$ so that $\sigma(t)$ is short-correlated. 

We conclude that $\sigma(t)$ is Gaussian, short-correlated random noise when $St\gtrsim 1$, $Fr\ll 1$. In the next Sections we use this result to find the spectrum of the Lyapunov exponents of inertial particles, cf. \cite{FFS}. 

\section{Lyapunov exponent of inertial particles} \label{spectrum}

The Lyapunov exponent $\lambda_1$ describes exponential growth of the distance $\bm r$ between two particles when $r\ll \eta$. If we designate the position of the first particle by $\bm x$ then $r\ll \eta$ implies that in the distance equation $\dot {\bm r}=\bm v(t, \bm x+\bm r)-\bm v(t, \bm x)$ we can use Taylor expansion that brings $\dot {\bm r}=\sigma \bm r$. Taking scalar product of the latter equation with $\bm r$ we find that the rate of change of the distance $r=|\bm r|$ is given by the component of $\sigma$ in the so-called major stretching direction $\bm {{\hat n}}=\bm r/r$,  
\begin{eqnarray}
&&\frac{d \ln r}{dt}=\bm {{\hat n}} \sigma \bm {{\hat n}},
\end{eqnarray}
where the name for $\bm {{\hat n}}$ comes from the observation that turbulence deforms balls into elongated ellipsoids whose major axis points in $\bm {{\hat n}}$ direction. The equation on $\bm {{\hat n}}$ results by using the above equation in $\dot {\bm r}=\dot r\bm {{\hat n}}+rd\bm {{\hat n}}/dt=\sigma\bm r$,
\begin{eqnarray}&&
\ \ \frac{d\bm {{\hat n}}}{dt}=\sigma\bm {{\hat n}}-\bm {{\hat n}}[\bm {{\hat n}} \sigma\bm {{\hat n}}].\label{orientation}
\end{eqnarray}
We find the solution  
\begin{eqnarray}
&&
\frac{1}{t}\ln\left(\frac{r(t)}{r(0)}\right)=\frac{1}{t}\int_0^t \bm {{\hat n}} \sigma\bm {{\hat n}}dt'. 
\end{eqnarray}
The law of large numbers implies that in the limit $t\to\infty$ the RHS tends to the constant given by the statistical average with probability one
\begin{eqnarray}
&&
\lambda_1=\lim_{t\to\infty}\frac{1}{t}\ln\left(\frac{r(t)}{r(0)}\right)=\lim_{t\to\infty} \langle \bm {{\hat n}}(t)\sigma(t)\bm {{\hat n}}(t)\rangle, \label{firstl}
\end{eqnarray}
where the limit $t\to\infty$ in the RHS is needed because it takes finite time for the statistics of $\bm {{\hat n}}(t)\sigma(t)\bm {{\hat n}}(t)$ to get stationary. To see this consider the average at $t=0$,
\begin{eqnarray}
&&
\langle \bm {{\hat n}}\sigma\bm {{\hat n}}\rangle(t=0)=\langle \sigma_{ik}(t=0)\rangle \langle \bm {{\hat n}}_i(t=0)\bm {{\hat n}}_k(t=0)\rangle=0,\nonumber
\end{eqnarray} 
where we used independence of $\bm {{\hat n}}(t=0)$ of $\sigma$ and that $\langle \sigma_{ik}(t=0)\rangle=\langle\nabla_kv_i(\bm x, 0)\rangle=0$ by spatial uniformity. In contrast $\langle \bm {{\hat n}}\sigma\bm {{\hat n}}\rangle(t)\neq 0$ when $t>0$ both because $\sigma$ and $\bm {{\hat n}}$ become dependent and because $\langle \sigma_{ik}(t)\rangle$ becomes non-vanishing. Consider the case when one can consider the statistics of $\sigma$ isotropic so that $\langle \sigma_{ik}\rangle=\delta_{ik}\langle tr\sigma\rangle/3$. Then the average is non-zero because preferential concentration of particles in regions with negative $\nabla\cdot\bm v=tr\sigma$ brings negative $\langle tr\sigma\rangle$ in the particle's frame (remind that $\sigma_{ik}(t)$ is the $\nabla_kv_i$ in the frame of the particle and not at fixed spatial position. The latter two frames coincide at $t=0$ but not later \cite{BFF,reviewp}).
 
We use the formula (\ref{firstl}) to calculate $\lambda_1$. Introducing the traceless component $\sigma'$ of $\sigma$ by $\sigma'_{ik}=\sigma_{ik}-tr \sigma \delta_{ik}/3$,
\begin{eqnarray}
&&\!\!\!\!\!\!\!\!
\lambda_1=\frac{\sum\lambda_i}{3}+\lim_{t\to\infty} \langle \bm {{\hat n}}(t)\sigma'(t)\bm {{\hat n}}(t)\rangle,\ \ \frac{d\bm {{\hat n}}}{dt}=\sigma'\bm {{\hat n}}-\bm {{\hat n}}[\bm {{\hat n}} \sigma'\bm {{\hat n}}],\nonumber
\end{eqnarray}
where we introduced the sum of Lyapunov exponents $\sum\lambda_i=\lim_{t\to\infty}\langle tr\sigma(t)\rangle$ that will be discussed later. To calculate $\lambda_1$ we use that $\sigma'$ is short-correlated in time employing the procedure that was introduced in \cite{FB}. 
We use the decomposition $\sigma'_{ik}=s'_{ik}-\epsilon_{ikl} w_l/2$ into the symmetric strain component $s'=(\sigma'+\sigma'^t)/2$ and the vorticity $\bm w=\nabla\times\bm v$ see e. g. \cite{Batchelor}, 
\begin{eqnarray}
&&\!\!\!\!\!\!\!\!\frac{d\bm {{\hat n}}}{dt}=\sigma'\bm {{\hat n}}-\bm {{\hat n}}[\bm {{\hat n}} \sigma'\bm {{\hat n}}]=s'\bm {{\hat n}}-\bm {{\hat n}}[\bm {{\hat n}} s'\bm {{\hat n}}]+\frac{\bm w\times\bm {{\hat n}}}{2},
\end{eqnarray}
where the last term corresponds to the well-known observation that $\bm w/2$ is the angular velocity of the local rotation of the fluid.
That term can be included by redefining the strain. We introduce $\bm {{\hat n}}=R\bm {{\hat n}}'$ where the orthogonal matrix $R$ obeys $\dot R_{ik}=\epsilon_{ilp}w_lR_{pk}/2$ with the initial condition $R(0)=1$. We find  
\begin{eqnarray}&&
\lambda_1=\frac{\sum\lambda_i}{3}+\lim_{t\to\infty} \langle \bm {{\hat n}}'(t){\tilde s}(t)\bm {{\hat n}}'(t)\rangle, \nonumber \\&&
\frac{d\bm {{\hat n}}'}{dt}={\tilde s}\bm {{\hat n}}'-\bm {{\hat n}}'[\bm {{\hat n}}' {\tilde s}\bm {{\hat n}}'], \nonumber
\end{eqnarray}
where ${\tilde s}=R^ts'R$ is symmetric. Below we omit prime in $\bm {{\hat n}}'$ with no ambiguity. We integrate the equation on $\bm {{\hat n}}$ from $t-\delta t$ to $t$ thus rewriting the differential equation on $\bm {{\hat n}}$ in integral form
 \begin{eqnarray}&&\!\!\!\!\!\!\!\!
\bm {{\hat n}}(t)=\bm {{\hat n}}_0+\int_{t-\delta t}^t \left({\tilde s}(t')\bm {{\hat n}}(t')-\bm {{\hat n}}(t')[\bm {{\hat n}}(t') {\tilde s}(t')\bm {{\hat n}}(t')]\right)dt'.\nonumber
 \end{eqnarray}
where $\bm {{\hat n}}_0=\bm {{\hat n}}(t-\delta t)$.
If $\delta t$ is much smaller than the inverse of the typical value $\sigma_c$ of $\sigma$ then the last term is small. The asymptotic series in $\sigma_c \delta t\ll 1$ is obtained by solving the equation by iterations. Below we will need terms of order up to $(\sigma_c \delta t)^2$. Neglecting higher order terms we have $\bm {{\hat n}}(t)=\bm {{\hat n}}_0+\bm {{\hat n}}_1+\bm {{\hat n}}_2$ where 
\begin{eqnarray}
&&\!\!\!\!\!\! \bm {{\hat n}}_{1i}=\int_{t-\delta t}^t dt'{\tilde s}_{kl}(t')\left[\delta_{ik}\bm {{\hat n}}_{0l}-\bm {{\hat n}}_{0i}\bm {{\hat n}}_{0k}\bm {{\hat n}}_{0l}\right],\nonumber\\&&\!\!\!\!\!\!
\bm {{\hat n}}_{2i}=\int_{t-\delta t}^t dt'({\tilde s}(t') \bm {{\hat n}}_1(t')- \bm {{\hat n}}_1(t')\left[\bm {{\hat n}}_0{\tilde s}(t')\bm {{\hat n}}_0\right]-2\bm {{\hat n}}_0
\nonumber\\&&\!\!\!\!\!\!
\left[\bm {{\hat n}}_0{\tilde s}(t')\bm {{\hat n}}_1(t')\right])_i\!=\!\int_{t-\delta t}^t dt'\int_{t-\delta t}^{t'} dt''{\tilde s}_{pn}(t'){\tilde s}_{kl}(t'')\nonumber\\&&\!\!\!\!\!\! 
(\delta_{ip}\delta_{nk}\bm {{\hat n}}_{0l}\!-\!\delta_{ip}\bm {{\hat n}}_{0n}\bm {{\hat n}}_{0k}\bm {{\hat n}}_{0l}
\!-\!\delta_{ik}\bm {{\hat n}}_{0n}\bm {{\hat n}}_{0p}\bm {{\hat n}}_{0l}\!
\nonumber\\&&\!\!\!\!\!\! 
-\!2\delta_{nk}\bm {{\hat n}}_{0i}\bm {{\hat n}}_{0p}\bm {{\hat n}}_{0l}
+3\bm {{\hat n}}_{0i}\bm {{\hat n}}_{0p}\bm {{\hat n}}_{0n}\bm {{\hat n}}_{0k}\bm {{\hat n}}_{0l})
.\label{it}
\end{eqnarray}
The Lyapunov exponent is determined by $\bm {{\hat n}}{\tilde s}\bm {{\hat n}}$ which in the considered order is given by 
\begin{eqnarray}
&& \bm {{\hat n}}(t){\tilde s}(t)\bm {{\hat n}}(t)=\bm {{\hat n}}_0{\tilde s}(t)\bm {{\hat n}}_0+2\bm {{\hat n}}_0{\tilde s}(t)\bm {{\hat n}}_1(t)
\nonumber\\&&
+\bm {{\hat n}}_1(t){\tilde s}(t)\bm {{\hat n}}_1(t)+2\bm {{\hat n}}_0{\tilde s}(t)\bm {{\hat n}}_2(t),
\end{eqnarray}
We impose the condition $\delta t\gg \tau$ which is possible due to $\sigma_c\tau\ll 1$.  Then $\bm {{\hat n}}_0$ which was determined by ${\tilde s}$ at times earlier than $t-\delta t$ can be considered independent of $\sigma'$ at time $t$. This is because $\delta t$ is much larger than the correlation time $\tau$ of $\sigma'(t)$. Thus in finding $\langle \bm {{\hat n}}(t){\tilde s}(t)\bm {{\hat n}}(t)\rangle$ we can average over the statistics of ${\tilde s}$ and $\bm {{\hat n}}_0$ independently. For instance $\langle\bm {{\hat n}}_0{\tilde s}(t)\bm {{\hat n}}_0\rangle=\langle{\tilde s}_{ik}\rangle\langle \bm {{\hat n}}_{0i}\bm {{\hat n}}_{0k}\rangle$ with correction which is exponentially small in the ratio of correlation time $\tau$ to $\delta t$. We find 
\begin{eqnarray}
&& \!\!\!\!\!\! \langle\bm {{\hat n}}(t){\tilde s}(t)\bm {{\hat n}}(t)\rangle\!-\!\langle{\tilde s}_{ik}\rangle\langle \bm {{\hat n}}_{0i}\bm {{\hat n}}_{0k}\rangle\!=\!2\langle\bm {{\hat n}}_0{\tilde s}(t)\bm {{\hat n}}_1(t)\rangle
\!+\!\langle\bm {{\hat n}}_1(t)
\nonumber\\&&\!\!\!\!\!\!
{\tilde s}(t)\bm {{\hat n}}_1(t)\rangle\!+\!2 \langle\bm {{\hat n}}_0{\tilde s}(t)\bm {{\hat n}}_2(t)\rangle\!=\!
2 \int_{t-\delta t}^t \!\!dt'
\langle{\tilde s}_{pi}(t){\tilde s}_{kl}(t')\rangle \nonumber\\&&\!\!\!\!\!\! 
\left[\delta_{ik}\langle \bm {{\hat n}}_{0p}\bm {{\hat n}}_{0l}\rangle-\langle \bm {{\hat n}}_{0p}\bm {{\hat n}}_{0i}\bm {{\hat n}}_{0k}\bm {{\hat n}}_{0l}\rangle\right]+2\int_{t-\delta t}^t dt'\int_{t-\delta t}^{t'} dt'' \nonumber\\&&\!\!\!\!\!\! \langle{\tilde s}_{ri}(t){\tilde s}_{pn}(t'){\tilde s}_{kl}(t'')\rangle(\langle \left[\delta_{rp}\bm {{\hat n}}_{0n}\!-\!\bm {{\hat n}}_{0r}\bm {{\hat n}}_{0p}\bm {{\hat n}}_{0n}\right][\delta_{ik}\bm {{\hat n}}_{0l}-\!\bm {{\hat n}}_{0i}
\nonumber\\&&\!\!\!\!\!\! 
\!\bm {{\hat n}}_{0k}\bm {{\hat n}}_{0l}]\rangle
\!+\!\langle (\delta_{ip}\delta_{nk}\bm {{\hat n}}_{0l}\bm {{\hat n}}_{0r}\!-\!\delta_{ip}\bm {{\hat n}}_{0n}\bm {{\hat n}}_{0k}\bm {{\hat n}}_{0l}\bm {{\hat n}}_{0r}\!-\!\delta_{ik}\bm {{\hat n}}_{0n}\bm {{\hat n}}_{0p}
\nonumber\\&&\!\!\!\!\!\!
\bm {{\hat n}}_{0l}\bm {{\hat n}}_{0r}\!-\!2\delta_{nk}\bm {{\hat n}}_{0i}\bm {{\hat n}}_{0p}\bm {{\hat n}}_{0l}\bm {{\hat n}}_{0r}
\!+\!3\bm {{\hat n}}_{0i}\bm {{\hat n}}_{0p}\bm {{\hat n}}_{0n}\bm {{\hat n}}_{0k}\bm {{\hat n}}_{0l}\bm {{\hat n}}_{0r})\rangle),\nonumber
\end{eqnarray}
where writing $\langle\bm {{\hat n}}_1(t){\tilde s}(t)\bm {{\hat n}}_1(t)\rangle$ we used the symmetry of the integrand in $t'$, $t''$ to pass to the integration domain $t''<t'$. When the small-scale turbulence can be considered isotropic we have $\langle \bm {{\hat n}}_{0i}\bm {{\hat n}}_{0k}\rangle=\delta_{ik}/3$ so that $\langle{\tilde s}_{ik}\rangle\langle \bm {{\hat n}}_{0i}\bm {{\hat n}}_{0k}\rangle\propto tr{\tilde s}=0$. Though the isotropy is presumed below the calculations can be performed in non-isotropic case (which is clear from the considerations so far) that is relevant to turbulence with not too large $Re$ or cases with shear. 
Using that the correlation time $\tau$ of $\sigma$ is much smaller than $\delta t$ we can write 
\begin{eqnarray}
&& \!\!\!\!\!\! \lambda_1=\frac{\sum \lambda_i}{3}
+\!2 \int_{-\infty}^t dt'\langle{\tilde s}_{pi}(t){\tilde s}_{kl}(t')\rangle [\delta_{ik}\langle \bm {{\hat n}}_{0p}\bm {{\hat n}}_{0l}\rangle
\nonumber\\&&\!\!\!\!\!\! -\langle \bm {{\hat n}}_{0p}\bm {{\hat n}}_{0i}\bm {{\hat n}}_{0k}\bm {{\hat n}}_{0l}\rangle]+2\int_{-\infty}^t dt'\int_{-\infty}^{t'} dt'' \langle{\tilde s}_{ri}(t){\tilde s}_{pn}(t')\nonumber\\&&\!\!\!\!\!\! {\tilde s}_{kl}(t'')\rangle\langle (\delta_{ik}\delta_{rp}\bm {{\hat n}}_{0n}\bm {{\hat n}}_{0l}+\delta_{ip}\delta_{nk}\bm {{\hat n}}_{0l}\bm {{\hat n}}_{0r}-\delta_{rp}\bm {{\hat n}}_{0n}\bm {{\hat n}}_{0i}\bm {{\hat n}}_{0k}
\nonumber\\&&\!\!\!\!\!\!\bm {{\hat n}}_{0l} -\!\delta_{ik}\bm {{\hat n}}_{0r}\bm {{\hat n}}_{0p}\bm {{\hat n}}_{0n}\bm {{\hat n}}_{0l}-\!\delta_{ip}\bm {{\hat n}}_{0n}\bm {{\hat n}}_{0k}\bm {{\hat n}}_{0l}\bm {{\hat n}}_{0r}-\!\delta_{ik}\bm {{\hat n}}_{0n}\bm {{\hat n}}_{0p}
\nonumber\\&&\!\!\!\!\!\!
\bm {{\hat n}}_{0l}\bm {{\hat n}}_{0r}\!-\!2\delta_{nk}\bm {{\hat n}}_{0i}\bm {{\hat n}}_{0p}\bm {{\hat n}}_{0l}\bm {{\hat n}}_{0r}
\!+\!4\bm {{\hat n}}_{0i}\bm {{\hat n}}_{0p}\bm {{\hat n}}_{0n}\bm {{\hat n}}_{0k}\bm {{\hat n}}_{0l}\bm {{\hat n}}_{0r})\rangle,\nonumber
\end{eqnarray}
Isotropy implies that $\bm {{\hat n}}$ is uniformly distributed over the unit sphere. Thus
\begin{eqnarray}
&&\!\!\!\!\!\! \langle \bm {{\hat n}}_{0i}\bm {{\hat n}}_{0k}\bm {{\hat n}}_{0l}\bm {{\hat n}}_{0m}\rangle=\frac{\delta_{ik}\delta_{lm}+\delta_{il}\delta_{km}+\delta_{im}\delta_{kl}}{15}.\label{quadr}
\end{eqnarray}
To find $\langle \bm {{\hat n}}_{0i}\bm {{\hat n}}_{0p}\bm {{\hat n}}_{0n}\bm {{\hat n}}_{0k}\bm {{\hat n}}_{0l}\bm {{\hat n}}_{0r}\rangle$ we write 
\begin{eqnarray}
&\langle \bm {{\hat n}}_{0i}\bm {{\hat n}}_{0p}\bm {{\hat n}}_{0n}\bm {{\hat n}}_{0k}\bm {{\hat n}}_{0l}\bm {{\hat n}}_{0r}\rangle=c\sum \delta_{i_1, i_2}\delta_{i_3, i_4}\delta_{i_5, i_6}.\label{sum}
\end{eqnarray}
where the sum runs over distinct pairs of indices and $c$ is a constant. Setting $p=i$ and summing over $i$ we find by comparing with Eq.~(\ref{quadr}) that $1/15=3c+4c$ where the first term on the RHS comes from terms where $(i, p)$ is one of the pairs in the sum in Eq.~(\ref{sum}) and the second from $12$ terms where there is no pair $(i, p)$. Thus $c=1/105$. We find 
\begin{eqnarray}
&& \!\!\!\!\!\! \lambda_1=\frac{\sum \lambda_i}{3}
+\!\frac{2}{5}\int_{-\infty}^t dt' \biggl[\langle tr{\tilde s}(t){\tilde s}(t')\rangle 
\nonumber\\&&
+\frac{3}{7}\int_{-\infty}^{t} dt''\langle tr{\tilde s}(t){\tilde s}(t'){\tilde s}(t'')\rangle 
\biggr], \label{rep}
\end{eqnarray}
where the limit $t\to\infty$ is implied and we used $tr{\tilde s}=0$. This coincides with the formula for incompressible case \cite{FB} up to $\sum \lambda_i/3$ term. One can write
\begin{eqnarray}
&&  \int_{-\infty}^t dt' \langle tr{\tilde s}(t){\tilde s}(t')\rangle 
\nonumber\\&&
=\int_{-\infty}^t dt' 
\langle tr R(t')R^t(t)s'(t)R(t)R^t(t')s'(t')\rangle.
\end{eqnarray}
Using the equation on $R$ to the lowest order one can use in the average $R_{ik}(t)-R_{ik}(t')\approx \epsilon_{ilp}R_{pk}(t)\int_{t'}^t w_l(t'')dt''/2$ where we use that $R$ varies over the same time-scale $1/\sigma_c$ as $\bm {{\hat n}}$. Multiplying both sides by $R_{sk}(t)$ we find that 
to lowest order $[R(t')R^t(t)]_{is}-\delta_{is}$ is given by antisymmetric matrix $a_{is}(t)=\epsilon_{isl}\int_{t'}^t w_l(t'')dt''/2$ so that $tr R(t')R^t(t)s'(t)R(t)R^t(t')s'(t')\approx tr s'(t)s'(t')+tr a(t)s'(t)s'(t')-tr s'(t)a(t)s'(t')$. We conclude that 
\begin{eqnarray}
&&  \!\!\!\!\!\!\int_{-\infty}^t dt' \langle tr{\tilde s}(t){\tilde s}(t')\rangle \approx \int_{-\infty}^t dt' \langle tr s'(t) s'(t')\rangle 
\nonumber\\&&\!\!\!\!\!\!
+\frac{\epsilon_{isl}}{2}\int_{-\infty}^t dt'\int_{t'}^tdt''\langle [s'(t)s'(t')-s'(t')s'(t)]_{si} w_l(t'')\rangle.\nonumber
\end{eqnarray}
Thus we obtain in the considered order that 
\begin{eqnarray}
&& \!\!\!\!\!\! \lambda_1=\frac{\sum \lambda_i}{3}
+\!\frac{2}{5}\int_{-\infty}^t dt'  \biggl[\langle tr s'(t) s'(t')\rangle +\frac{\epsilon_{isl}}{2}\int_{-\infty}^t dt'
\nonumber\\&&
\int_{t'}^tdt''\langle [s'(t)s'(t')-s'(t')s'(t)]_{si} w_l(t'')\rangle
\nonumber\\&&
+\frac{3}{7}\int_{-\infty}^{t} dt''\langle tr s'(t)s'(t')s'(t'')\rangle 
\biggr],
\end{eqnarray}
where we observed in the last line that in the considered order one can put $s'$ instead of ${\tilde s}$. 
The calculation of $\sum\lambda_i$ performed in the next Section demonstrates that $\sum\lambda_i/\lambda_1\propto Fr$ so that the first term in $\lambda_1$ can be dropped. We find
\begin{eqnarray}
&& \!\!\!\! \lambda_1\approx \frac{2}{5}\int_{-\infty}^t  \langle tr s'(t) s'(t')\rangle dt',\label{rep0}
\end{eqnarray}
cf. \cite{FB}. Finally to leading order in weak compressibility we can use the symmetric component $\sigma_l^s=(\sigma_l+\sigma_l^t)/2$ of traceless matrix $\sigma_l$ as $s'$, 
\begin{eqnarray}
&& \lambda_1\approx \frac{2}{5}\int_{-\infty}^t  \langle tr \sigma_l^s(t) \sigma_l^s(t')\rangle dt'.
\end{eqnarray}
To leading order in weak compressibility, writing $\sigma_l^s$ in components one can write  
\begin{eqnarray}&&\!\!\!\!\!\!\!\!\!\!\!\!\!\!
\lambda_1 =\frac{1}{5} \int_{-\infty}^0 \left\langle (\sigma_l)_{ik}[0, 0] (\sigma_l)_{ik}[t, \bm X(t, 0)] \right\rangle dt\nonumber\\&&
+\frac{1}{5} \int_{-\infty}^0 \left\langle (\sigma_l)_{ki}[0, 0] (\sigma_l)_{ik}[t, \bm X(t, 0)] \right\rangle dt, \label{first}
\end{eqnarray}
where $\bm X(t, \bm x)$ are the trajectories of the incompressible component of the flow $\bm v_{\perp}$ obeying 
\begin{eqnarray}
&&\partial_t\bm X(t, \bm x)=\bm v_{\perp}\left[t, \bm X(t, \bm x)\right],\ \ \bm X(0, \bm x)=\bm x.
\end{eqnarray}
To the same order we use 
\begin{eqnarray}&&\!\!\!\!\!\!\!\!\!\!\!\!
\sigma_l[t, \bm X(t, 0)]\approx \int_{-\infty}^t \exp\left[\frac{t'-t}{\tau}\right]s[t', \bm X(t', 0)]\frac{dt'}{\tau},
\end{eqnarray}
where we put in the RHS the trajectory of $\bm v_{\perp}$ and not the complete $\bm v$. Due to incompressibility of $\bm v_{\perp}$ the correlation functions in the integrand of Eq.~(\ref{first}) are even functions of $t$ so that we can write 
\begin{eqnarray}&&\!\!\!\!\!\!\!\!\!\!\!\!\!\!
\lambda_1 =\frac{1}{10} \int_{-\infty}^{\infty} \left\langle (\sigma_l)_{ik}[0, 0] (\sigma_l)_{ik}[t, \bm X(t, 0)] \right\rangle dt\nonumber\\&&
+\frac{1}{10} \int_{-\infty}^{\infty} \left\langle (\sigma_l)_{ki}[0, 0] (\sigma_l)_{ik}[t, \bm X(t, 0)] \right\rangle dt, \label{first}
\end{eqnarray}
Using Eq.~(\ref{linear}) we find 
\begin{eqnarray}&&
\lambda_1 =\int_{-\infty}^0dt_1\int_{-\infty}^{\infty} dt\int_{-\infty}^t\frac{dt_2 }{10\tau^2}  
\exp\left[\frac{t_1+t_2-t}{\tau}\right] 
\nonumber\\&&
(\langle s_{ik}[t_1, \bm X(t_1, 0)] s_{ik}[t_2, \bm X(t_2, 0) \rangle
\nonumber\\&&
+\langle s_{ki}[t_1, \bm X(t_1, 0)] s_{ik}[t_2, \bm X(t_2, 0)]\rangle) .\label{ngl}
\end{eqnarray}
Interchanging the order of integrations over $t$ and $t_2$ we can integrate over $t$, 
\begin{eqnarray}&&\!\!\!\!\!\!\!\!
\lambda_1\!=\!\int_{-\infty}^0dt_1\!\int_{-\infty}^{\infty}\!\frac{dt_2 }{10\tau}  
(\langle s_{ik}[t_1, \bm X(t_1, 0)] s_{ik}[t_2, \bm X(t_2, 0) \rangle
\nonumber\\&&\!\!\!\!\!\!\!\!
+\langle s_{ki}[t_1, \bm X(t_1, 0)] s_{ik}[t_2, \bm X(t_2, 0)]\rangle)\exp\left[\frac{t_1}{\tau}\right].
\end{eqnarray}
Observing that the integral over $t_2$ is independent of $t_1$ due to stationarity we can calculate the integral over $t_1$,
\begin{eqnarray}&&
\lambda_1 =\int_{-\infty}^{\infty}\frac{dt}{10}  
(\langle s_{ik}(0, 0) s_{ik}[t, \bm X(t, 0)] \rangle
\nonumber\\&&
+\langle s_{ki}(0, 0)] s_{ik}[t, \bm X(t, 0)]\rangle).
\end{eqnarray}
We observe that since the temporal correlations of $s_{ik}[t, \bm X(t, 0)]$ are determined by gravitational drift through "frozen flow" then (see proof below) 
\begin{eqnarray}&&
\lambda_1= \frac{1}{10}
\int_{-\infty}^{\infty}dt( \left\langle \nabla_{k}u_i(0) \nabla_{k}u_i(\bm g\tau t) \right\rangle 
\nonumber\\&&
+\left\langle \nabla_{k}u_i(0) \nabla_{i}u_k(\bm g\tau t) \right\rangle. \label{formula}
\end{eqnarray}
The last term vanishes due to incompressibility which is seen by integrating by parts in the spatial average, $\left\langle \nabla_{k}u_i(0) \nabla_{i}u_k(\bm g\tau t) \right\rangle=-\langle u_i(0) \nabla_i\nabla_{k}u_k(\bm g\tau t) \rangle=0$. Finally using isotropy of the small-scale statistics of turbulence we find
\begin{eqnarray}&&
\lambda_1=\frac{1}{10g\tau}\int_{-\infty}^{\infty} \left\langle \nabla_{k}u_i(0) \nabla_{k}u_i(x) \right\rangle dx ,
\label{formula1}
\end{eqnarray}
Isotropy holds at the Kolmogorov scale in typical situations including our numerical simulations so it is presumed below.

Formula (\ref{formula}) can be proved by noting that the separation of inertial and fluid particles is ballistic, $\bm X(t, 0)=\bm x^u(t, 0)+\bm g t\tau$ at relevant time-scales $\tau_g\ll t_{\eta}$. Here $\bm x^u(t, 0)$ is the Lagrangian trajectory of the fluid particle defined by Eq.~(\ref{Lagrangian}). 
To demonstrate the ballistic separation we observe that $\bm r(t)=\bm X(t, 0)-\bm x^u(t, 0)$ obeys 
\begin{eqnarray}&&\!\!\!\!\!\!\!\!\!\!\!\!\!\!
\dot {\bm r}=\bm v[t, \bm x^u(t, 0)+\bm r]-\bm u[t, \bm x^u(t, 0)]=\sigma \bm r+\bm v_{d},\ \ \label{rate}
\end{eqnarray}
where the drift velocity $\bm v_{d}=\bm v[t, \bm x^u(t, 0)]-\bm u[t, \bm x^u(t, 0)]$ obeys $\bm v_{d}=\bm v_g$ in the considered range of parameters (the equation (\ref{rate}) holds at not too large $t$ where $r(t)\ll \eta$). It follows that at times $t$ where $r\ll v_g/\sigma_c$ one can neglect $\sigma\bm r$ in the last term in Eq.~(\ref{rate}) producing $\dot {\bm r}=\bm v_{d}$ and $\bm r(t)=\bm v_g t$. The demand of self-consistency limits the time of applicability of $\bm r(t)=\bm v_g t$ to $t\ll 1/\sigma_c$. Using the characteristic value $\sigma_c\sim Fr^{1/2}/\tau$ we find that at $t\ll \tau_g St^2 Fr^{-3/2}$ one has $\bm r(t)=\bm v_g t$. Since  $\tau_g St^2 Fr^{-3/2}$ is much larger than the correlation time $\tau_g$ of $s$ then it follows that 
\begin{eqnarray}&& \!\!\!\!\!\!
\int\left\langle s_{ik}(0, 0) s_{ik}[t, \bm X(t, 0)] \right\rangle dt
=\int dt \langle s_{ik}(0, 0)
\nonumber\\&&\!\!\!\!\!\!
\times s_{ik}[t, \bm x^u(t, 0)+\bm v_g t] \rangle 
\approx \int \left\langle s_{ik}(0) s_{ik}(\bm v_g t) \right\rangle dt, 
\end{eqnarray}
where we noticed that at time-scales of order $\tau_g\ll \tau_{\eta}$ one has $s_{ik}[t, \bm X^u(t, 0) + \bm v_g t] \approx s_{ik}[0, \bm v_g t]$ since at times much smaller than $t_{\eta}$ the field $\bm u$ is ``rigidly" transferred in the Lagrangian frame.  This reproduces Eq.~(\ref{formula}).

It is possible to rewrite $\lambda_1$ in terms of the spectrum of turbulence $E(k)$ by observing that
\begin{eqnarray}&&\!\!\!\!\!\!\!
\langle \nabla_ku_i(0) \nabla_ku_i(\bm x)\rangle=-\nabla^2 \langle u_i(0) u_i(\bm x)\rangle
\nonumber\\&&\!\!\!\!\!\!\!
=\!-\!\nabla^2\int E(k)\exp[i\bm k\cdot \bm x]\frac{d\bm k}{2\pi k^2}\!=\!
\int_0^{\infty} \frac{2E(k)\sin(kx)kdk}{x}.\nonumber
\end{eqnarray} 
Using Eq.~(\ref{formula1}) we find 
\begin{eqnarray}&&\!\!\!\!\!\!\!
\lambda_1\!=\!\int_0^{\infty} E(k)kdk\int dx \frac{\sin(kx)}{5 g\tau x}\!=\!\frac{\pi\int_0^{\infty} E(k)kdk}{5 g\tau}. 
\end{eqnarray}
We conclude that $\lambda_1\tau$ is independent of the properties of particles (measured by the Stokes number), 
\begin{eqnarray}&&\!\!\!\!\!\!\!
\lambda_1\tau=\frac{\pi\int_0^{\infty} E(k)kdk}{5 g}\propto Fr. 
\end{eqnarray}
Thus we wrote the Lyapunov exponent in terms of the spectrum that characterizes instantaneous statistics of turbulence rather than different time statistics that determines the Lyapunov exponent of passive tracers.
This fits the physics of the description: the particles' drift through the flow makes them pass many correlation lengths $\eta$ of the turbulent gradients during their relaxation time $\tau$. As a result, the particles react to the accumulated action of lots of independent turbulent vortices (understood here as configurations of velocity gradient with characteristic correlation length $\eta$) seeing turbulence as frozen Gaussian field
which is completely characterized by the power - integral pair-correlation function of gradients "seen" by the particle falling at the speed $\bm g\tau$, cf. Eq.~(\ref{formula}). 

We observe that the calculation of $\lambda_1$ performed in this Section relied on the possibility to introduce the flow and the short-correlatedness of $\sigma$. In fact, both properties hold in the wider range $St\gg Fr$ (remind that $Fr\ll 1$ is presumed). Indeed we saw that both properties hold when $St\gtrsim 1$ so we only need to consider $St\ll 1$. Then if $\tau\ll \min[\tau_\eta, \tau_g]$ then the particles' flow is given by Eq.~(\ref{eff_vel}).
If $St\ll Fr$ or $\tau_g\gg \tau_{\eta}$ then gravity influences negligibly velocity gradients in the particle's frame. 
In that case $\lambda_1$ is close to the Lyapunov exponent of the fluid particles $\lambda_1^{turb}$ because without gravity the effect of inertia on $\lambda_1$ is small when $St\ll 1$. In contrast if $St^2\ll Fr\ll St$ or $\tau\ll \tau_g\ll \tau_{\eta}$ then the correlation time of particles' flow gradients is $\tau_g$ which product with the typical value of velocity gradients $1/\tau_{\eta}$ is small \cite{FFS,FHa}. Finally in the case $\tau\gtrsim \tau_g$ or $St^2\gtrsim Fr$ though Eq.~(\ref{eff_vel}) breaks down one can still introduce the particles' flow because the condition (\ref{condition}) holds at $St\ll 1$ independently of $Fr$. Then the smallness of the product of the correlation time $\tau$ of velocity gradients and the typical value of the gradients $\sigma$ follows from Eq.~(\ref{condition}) because smallness of $\sigma\tau$ is the condition that Eq.~(\ref{vel_gradients2}) does not produce finite time blow-up. We conclude that 
\begin{eqnarray}&&\!\!\!\!\!\!\!
\lambda_1\tau=\frac{\pi\int_0^{\infty} E(k)kdk}{5 g}\propto Fr,\ \ St\gg Fr. \label{finalf}
\end{eqnarray}
Together with the observation that $\lambda_1$ is given by $\lambda_1^{turb}$ when $St\ll Fr$ this provides complete description of the Lyapunov exponent of inertial particles when $Fr\ll 1$.

\section{Third Lyapunov exponent and smoothness scale of particles' flow}\label{third}

The calculation of the previous Section can be used to find readily the third Lyapunov exponent of particles $\lambda_3$. That describes the contraction of the smallest axis of the ellipsoid into which turbulence deform small balls of particles. Reversing the time direction one concludes that $\lambda_3$ determines the exponential divergence of particles' trajectories backward in time (this statement demands a correction due to finite compressibility see \cite{BFF}. That correction however can be neglected in our case due to smallness of compressibility). 

It was demonstrated in \cite{FB} that a representation that is very close to Eqs.~(\ref{orientation}), (\ref{firstl}) for $\lambda_1$  holds for $\lambda_3$, 
\begin{eqnarray}
&&\!\!\!\!\!\!\!\!\!\!\!\!\!\!\!\!
\lambda_3\!=\!\lim_{t\to\infty} \langle \bm {{\hat n'}}(t)\sigma(t)\bm {{\hat n'}}(t)\rangle,\ \ \frac{d\bm {{\hat n'}}}{dt}\!=\!-\sigma^t\bm {{\hat n}}\!+\!\bm {{\hat n'}}[\bm {{\hat n'}} \sigma\bm {{\hat n'}}]. \label{thirdl}
\end{eqnarray}
Thus using $-\sigma^t$ instead of $\sigma$ in the calculation of $\lambda_1$ one finds $-\lambda_3$. Since the change $\sigma\to-\sigma^t$ does not change $\bm w$ but reverses the sign of $s'$ then the formula for $\lambda_3$ is obtained by changing ${\tilde s}$ to $-{\tilde s}$ in Eq.~(\ref{rep}) which gives
\begin{eqnarray}
&& \!\!\!\!\!\! \lambda_3=\frac{\sum \lambda_i}{3}
-\!\frac{2}{5}\int_{-\infty}^t dt' \biggl[\langle tr{\tilde s}(t){\tilde s}(t')\rangle 
\nonumber\\&&
-\frac{3}{7}\int_{-\infty}^{t} dt''\langle tr{\tilde s}(t){\tilde s}(t'){\tilde s}(t'')\rangle 
\biggr], \label{rep3}
\end{eqnarray}
To leading order the second term dominates the formula,
\begin{eqnarray}
&& \lambda_3\approx 
-\frac{2}{5}\int_{-\infty}^t dt' \langle tr{\tilde s}(t){\tilde s}(t')\rangle. 
\end{eqnarray}
Comparing with Eq.~(\ref{rep0}) we conclude
\begin{eqnarray}
&&\!\!\!\!\!\!\!\!\!\!\!\! \lambda_3\approx -\lambda_1=-\frac{\pi\int_0^{\infty} E(k)kdk}{5 g\tau},\ \ Fr\ll \min[1, St],
\end{eqnarray}
cf. \cite{FB}. 
This result can be used to demonstrate that when $Fr\ll \min[1, St]$ the scale of smoothness of $\bm v(t, \bm x)$ is the same Kolmogorov scale $\eta$ as for turbulence itself. This is not obvious because the dependence of $\bm v(t, \bm x)$ on $\bm u(t, \bm x)$ is non-local in time. Thus comparing the gradients $\sigma$ of $\bm v(t, \bm x)$ at nearby points 
\begin{eqnarray}&&\!\!\!\!\!\!\!\!\!\!\!\!
\sigma(0, \bm x_i)\approx\int_{-\infty}^0 \exp\left[\frac{t}{\tau}\right]s[t, \bm X(t, \bm x_i)]\frac{dt}{\tau},\label{linear1}
\end{eqnarray}
where $i=1, 2$, one can have the variation of $\sigma$ caused by the exponential divergence of the trajectories $\bm X(t, \bm x_i)$ back in time. This divergence occurs at the exponent $|\lambda_3|=\lambda_1$ that we demonstrated to obey $|\lambda_3|\tau\sim Fr\ll 1$ though. Correspondingly for times of order $\tau$ relevant to the integral in Eq. (\ref{linear1}) one has $\bm X(t, \bm x_1)-\bm X(t, \bm x_2)\approx \bm x_1-\bm x_2$ for $|\bm x_1-\bm x_2|\ll \eta$. It follows that 
$s[t, \bm X(t, \bm x_1)]\approx s[t, \bm X(t, \bm x_2)]$ at relevant times in Eq. (\ref{linear1}) implying 
\begin{eqnarray}&&\!\!\!\!\!\!\!\!\!\!\!\!
\sigma(0, \bm x_1)\approx \sigma(0, \bm x_2), \ \ |\bm x_1-\bm x_2|\ll \eta.
\end{eqnarray} 
On the contrary if $|\bm x_1-\bm x_2|\sim \eta$ then the difference of $\sigma(0, \bm x_i)$ is of order one. Correspondingly the spatial scale of variations of gradients of $\bm v$ is $\eta$ when $Fr\ll \min[1, St]$. 

In the remaining range $St\ll 1$, $Fr\gtrsim St$ where the flow can be defined one has $\tau\ll \min[\tau_{\eta}, \tau_g]$ so that $\bm v$ is given by Eq.~(\ref{eff_vel}) that implies smoothness at the scale $\eta$. We conclude that $\bm v(t, \bm x)$ has the same scale of smoothness $\eta$ as turbulence when it can be defined.  

\section{Rate of contraction of particles' volumes when $St\gtrsim 1$, $Fr\ll 1$} \label{sumsection}

In this Section we consider the key consequence of finite compressibility of the flow of particles - the rate of exponential contraction of infinitesimal volumes $V$. The latter obey $\dot V=tr\sigma V$, see e. g. \cite{Batchelor} (the rate of change of infinitesimal volume is the flow integral over its surface $\int \bm v\cdot \bm dS=\int \nabla\cdot \bm v dV$ which for small volumes reduces to $\nabla\cdot\bm v V$ where $\nabla\cdot\bm v =tr\sigma$). Thus $\ln [V(t)/V(0)]=\int_0^t tr\sigma(t') dt'$ so that
\begin{eqnarray}
&&
\lim_{t\to\infty}\frac{1}{t}\ln\left(\frac{V(t)}{V(0)}\right)=\lim_{t\to\infty} \frac{1}{t} \int_0^t tr\sigma(t')dt'.\label{vol1}
\end{eqnarray}
The logarithmic rate of growth of infinitesimal volumes defines the sum of the Lyapunov exponents $\sum\lambda_i$
\begin{eqnarray}
&&
\sum\lambda_i=\lim_{t\to\infty}\frac{1}{t}\langle tr\sigma(t)\rangle,\label{vol2}
\end{eqnarray}
cf. \cite{reviewp} and references therein. To find $\sum\lambda_i$ it is simpler to use (\ref{sumlambda}) with $\nabla\cdot \bm v=tr \sigma$ rather than to try to proceed like in the calculation of $\lambda_1$ (which in fact brings to the same formula, see \cite{BFF}). To leading order in weak compressibility one can consider $\langle tr \sigma(0)tr\sigma(t)\rangle$ to be even function of $t$ because to that order one could use the trajectories of the incompressible flow in the spatial arguments of $\nabla\cdot\bm v$ (which produce even correlation functions) instead of the true trajectories of the particles. Thus we use 
\begin{eqnarray}&&
\sum \lambda_i\approx-\frac{1}{2}\int_{-\infty}^{\infty} \langle tr \sigma(0)tr\sigma(t)\rangle dt,
\end{eqnarray} 
that is neater in calculations then Eq.~(\ref{sumlambda}).  Using Eq.~(\ref{trace}) we obtain
\begin{eqnarray}&&
\sum \lambda_i\approx -\frac{1}{2}\int_{-\infty}^{\infty}  dt \int_{-\infty}^0 dt_1\int_{-\infty}^t dt_2\exp\left[\frac{t_1+t_2-t}{\tau}\right]
\nonumber\\&&
\langle tr \sigma_l^2(t_1)tr \sigma_l^2(t_2)\rangle. \label{sum}
\end{eqnarray} 
Interchanging the order of integrations over $t$ and $t_2$ and integrating over $t$ we find 
\begin{eqnarray}&&\!\!\!\!\!\!\!
\sum \lambda_i\!\approx\! -\frac{\tau}{2}\int_{-\infty}^{\infty}\!\!\!\!  dt_2 \!\int_{-\infty}^0\! \!\!\!\!dt_1\exp\left(\frac{t_1}{\tau}\right)
\langle tr \sigma_l^2(t_1)tr \sigma_l^2(t_2)\rangle. \nonumber
\end{eqnarray}
In the leading order one has $(\sigma_l)_{ik}(t)=\nabla_kv_i[t, \bm X(t, 0)]$ in the formula above. The correlation function $\langle tr \sigma_l^2(t_1)tr \sigma_l^2(t_2)\rangle$ depends on the difference $|t_2-t_1|$ only due to incompressibility of $\bm v_{\perp}$ that determines $\bm X(t, 0)$ and stationarity. Integrating over $t_2$ one finds $t_1-$independent result producing 
\begin{eqnarray}&&\!\!\!\!\!\!\!
\sum \lambda_i\!\approx\! -\frac{\tau^2}{2}\int_{-\infty}^{\infty} 
\langle tr \sigma_l^2(0)tr \sigma_l^2(t)\rangle  dt. 
\end{eqnarray}
This formula is the counterpart of Eq.~(\ref{first}) for $\lambda_1$. We recall that $\sigma_l(t)$ is Gaussian process so that its correlation functions can be obtained using the Wick's theorem 
\begin{eqnarray}&&
\langle tr \sigma_l^2(0)tr \sigma_l^2(t)\rangle=\langle tr \sigma_l^2(0)\rangle \langle  tr \sigma_l^2(t)\rangle
\nonumber\\&&
+2\langle  (\sigma_l)_{ik}(0)(\sigma_l)_{pr}(t)\rangle\langle(\sigma_l)_{ki}(0)(\sigma_l)_{rp}(t)\rangle.
\end{eqnarray}
We observe that in this order the first term on the RHS can be neglected. We have $\langle tr \sigma_l^2(t_1)\rangle=\langle \nabla\cdot\left[(\bm v\cdot \nabla)\bm v\right]\rangle-\langle \left[(\bm v\cdot \nabla)\nabla\cdot\bm v\right]\rangle=\langle \left[\nabla\cdot\bm v\right]^2\rangle$ where in spatial averaging we integrated by parts and neglected the contribution of the boundaries. It follows that $\langle tr \sigma_l^2(t_1)\rangle \langle  tr \sigma_l^2(t_2)\rangle$ is of the fourth order in the compressible component of velocity and can be neglected. We obtain 
\begin{eqnarray}&&\!\!\!\!\!\!\!
\sum \lambda_i\!\approx\! -\tau^2\!\int_{-\infty}^{\infty}\!\!\!\! \!dt \langle  (\sigma_l)_{ik}(0)(\sigma_l)_{pr}(t)\rangle\langle(\sigma_l)_{ki}(0)(\sigma_l)_{rp}(t)\rangle\nonumber\\&&\!\!\!\!\!\!\!=
- \int_{-\infty}^{\infty} \frac{dt}{\tau^2}\int_{-\infty}^t dt_2dt_4 \int_{-\infty}^0 dt_1dt_3\langle 
s_{ik}(t_1)s_{pr}(t_2)\rangle\nonumber\\&&\!\!\!\!\!\!\!\langle s_{ki}(t_3)s_{rp}(t_4)\rangle\exp\left[\frac{t_1+t_2+t_3+t_4-2t}{\tau}\right],
\end{eqnarray}
where we used Eq.~(\ref{linear}).
We observe that the integral over $t_1$, $t_3$ to the right of $dt_2dt_4$ produces symmetric function of $t_2$ and $t_4$ so we can write 
\begin{eqnarray}&&\!\!\!\!\!\!\!
\sum \lambda_i\!\approx\!
- \int_{-\infty}^{\infty} \!\frac{2dt}{\tau^2}\!\int_{-\infty}^t \!\!dt_2\int_{-\infty}^{t_2}\!
dt_4 \!\int_{-\infty}^0 \!\!dt_1\langle 
s_{ik}(t_1)s_{pr}(t_2)\rangle\nonumber\\&&\!\!\!\!\!\!\!\int_{-\infty}^0 dt_3 
\langle s_{ki}(t_3)s_{rp}(t_4)\rangle\exp\left[\frac{t_1+t_2+t_3+t_4-2t}{\tau}\right].
\end{eqnarray}
Interchanging the order of integrations over $t$, $t_2$, $t_4$ and performing the integral over $t$ we find 
\begin{eqnarray}&&\!\!\!\!\!\!\!
\sum \lambda_i\!\approx\!
-\frac{2}{\tau^2} \int_{-\infty}^{\infty} dt_4\int_{t_4}^{\infty} dt_2\int_{t_2}^{\infty}
dt \int_{-\infty}^0 dt_1dt_3\langle 
s_{ik}(t_1)\nonumber\\&&\!\!\!\!\!\!\!\times s_{pr}(t_2)\rangle \langle s_{ki}(t_3)s_{rp}(t_4)\rangle\exp\left[\frac{t_1+t_2+t_3+t_4-2t}{\tau}\right]\nonumber\\&&\!\!\!\!\!\!\!=-\frac{1}{\tau} \int_{-\infty}^{\infty} dt_4\int_{t_4}^{\infty} dt_2 \int_{-\infty}^0 dt_1dt_3\langle 
s_{ik}(t_1)s_{pr}(t_2)\rangle\nonumber\\&&\!\!\!\!\!\!\! \langle s_{ki}(t_3)s_{rp}(t_4)\rangle\exp\left[\frac{t_1-t_2+t_3+t_4}{\tau}\right].
\end{eqnarray}
Further we observe that since $\langle s_{ik}(t_1)s_{pr}(t_2)\rangle$ decays fast when $|t_2-t_1|$ exceeds $\tau_g\ll \tau$ then we can neglect $t_1-t_2$ term  in the exponent and set similarly $t_4=t_3$,
\begin{eqnarray}&&\!\!\!\!\!\!\!
\sum \lambda_i\!=-\frac{1}{\tau} \int_{-\infty}^{\infty} dt_4\int_{t_4}^{\infty} dt_2 \int_{-\infty}^0 dt_1dt_3\langle 
s_{ik}(t_1)s_{pr}(t_2)\rangle\nonumber\\&&\!\!\!\!\!\!\! \langle s_{ki}(t_3)s_{rp}(t_4)\rangle\exp\left[\frac{2t_4}{\tau}\right].
\end{eqnarray}
Finally we observe that $\int_{-\infty}^0 dt_1 \langle s_{ik}(t_1)s_{pr}(t_2)\rangle$ is a function of $t_2$ that decreases from constant value $C_{ikpr}=\int_{-\infty}^{\infty} dt \langle s_{ik}(t)s_{pr}(0)\rangle$ at negative $t_2$ to zero at positive $t_2$ within the vicinity of $t_2=0$ of order $\tau_g$. Since this vicinity is much smaller than the effective domain of integration of order $\tau$ then we put $\int_{-\infty}^0 dt_1 \langle s_{ik}(t_1)s_{pr}(t_2)\rangle=C_{ikpr}\theta(-t_2)$ where $\theta(x)$ is the step function. Using similar consideration for $\int_{-\infty}^0 dt_3\langle s_{ki}(t_3)s_{rp}(t_4)\rangle$ we find 
\begin{eqnarray}&&\!\!\!\!\!\!\!
\sum \lambda_i\!=-\frac{C_{ikpr}C_{kirp}}{\tau} \int_{-\infty}^{0} dt_4\int_{t_4}^{0} dt_2 \exp\left[\frac{2t_4}{\tau}\right]\nonumber\\&&
=-\frac{\tau C_{ikpr}C_{kirp}}{4}.\label{sum1}
\end{eqnarray}
Following the same considerations that were used in the study of $\lambda_1$ we can write 
\begin{eqnarray}&&\!\!\!\!\!\!\!
C_{ikpr}=\int_{-\infty}^{\infty} \langle\nabla_r u_p(0)\nabla_ku_i(\bm g\tau t)\rangle dt=\frac{c_{ikpr}}{g\tau},\label{sum2}
\end{eqnarray}
where the tensor $c_{ikpr}$ is independent of the properties of the particles but rather is a property of turbulence,
\begin{eqnarray}&&\!\!\!\!\!\!\!
c_{ikpr}({\hat z})=\int  \langle \nabla_ru_p(0) \nabla_ku_i(t{\hat z})\rangle dt.\label{defc}
\end{eqnarray}
This tensor coincides up to multiplicative constant with $F_{ikpr}(\bm w)$ introduced in \cite{FH} (this is seen using $\langle\nabla_r u_p(0)\nabla_ku_i(t{\hat z})\rangle=\langle\nabla_r u_p(-t{\hat z})\nabla_ku_i(0)\rangle$ and changing the sign of the integration variable).  
To find $c_{ikpr}({\hat z})$ we observe that (${\hat k}=\bm k/k$)
\begin{eqnarray}&& 
\langle \nabla_ru_p(0) \nabla_ku_i(\bm x)\rangle=-\nabla_r\nabla_k\langle u_p(0) u_i(\bm x)\rangle
\nonumber\\&&
=-\nabla_r\nabla_k
\int E(k)\left[\delta_{ip}-{\hat k}_i{\hat k}_p\right]\exp[i\bm k\cdot \bm x]\frac{d\bm k}{4\pi k^2}
\nonumber\\&&
=\int E(k)k_rk_k\left[\delta_{ip}-{\hat k}_i{\hat k}_p\right]\exp[i\bm k\cdot \bm x]\frac{d\bm k}{4\pi k^2}.
\end{eqnarray}
The use of isotropic form of the spectrum which does not hold at small wave-numbers is not a limitation since the integrals are determined by wave-numbers of the order of $\eta^{-1}$ for which isotropy holds well. Using Eq.~(\ref{defc}) we find
\begin{eqnarray}&&\!\!\!\!\!\!\!
c_{ikpr}({\hat z})=\int dt\int E(k)k_rk_k\left[\delta_{ip}-{\hat k}_i{\hat k}_p\right]\exp[ik_zt]\frac{d\bm k}{4\pi k^2}
\nonumber\\&&\!\!\!\!\!\!\!
=\int E(k){\hat k}_r{\hat k}_k\delta(k_z)\left[\delta_{ip}-{\hat k}_i{\hat k}_p\right]\frac{d\bm k}{2}.
\end{eqnarray}
We conclude that 
\begin{eqnarray}&&\!\!\!\!\!\!\!
c_{ikpr}({\hat z})c_{kirp}({\hat z})
=\int \frac{E(k)E(k')\delta(k_z)\delta(k'_z)d\bm kd\bm k'}{4}
\nonumber\\&&\!\!\!\!\!\!\!
\left[\delta_{ip}{\hat k}_r{\hat k}_k-{\hat k}_i{\hat k}_p{\hat k}_r{\hat k}_k\right]
\left[\delta_{kr}{\hat k'}_p{\hat k'}_i-{\hat k'}_k{\hat k'}_r{\hat k'}_p{\hat k'}_i\right]\nonumber\\&&\!\!\!\!\!\!\!
=\int \frac{E(k)E(k')\delta(k_z)\delta(k'_z)d\bm kd\bm k'}{4}
\left[1-({\hat k}\cdot{\hat k}')^2\right]^2.
\end{eqnarray}
The $\delta-$functions turn the integrals into two-dimensional ones. Taking $\bm k$ in the direction of $x-$axis we have 
\begin{eqnarray}&&\!\!\!\!\!\!\!
c_{ikpr}({\hat z})c_{kirp}({\hat z})
=\int \frac{E(k)d\bm k}{4}\int E(k')k'dk'd\phi\sin^4\phi\nonumber\\&&
=\frac{3[\pi\int E(k)kdk]^2}{8}.
\end{eqnarray}
We conclude using Eqs.~(\ref{sum1})-(\ref{sum2}) that 
\begin{eqnarray}&&\!\!\!\!\!\!\!
\sum \lambda_i\tau=-\frac{3[\pi\int E(k)kdk]^2}{32g^2}\propto Fr^2.\label{finalsum}
\end{eqnarray}
Thus we find that $\sum \lambda_i\tau$ is a negative number which is independent of the properties of the particles and proportional to $Fr^2$. The non-positivity of $\sum \lambda_i$ is the universal property that can be traced to the second law of thermodynamics \cite{FF} but its strict negativity and magnitude are non-trivial. We conclude that volumes of particles decrease exponentially at the rate proportional to $Fr^2/\tau$. 

\begin{figure}
\includegraphics[width=8.0 cm,clip=]{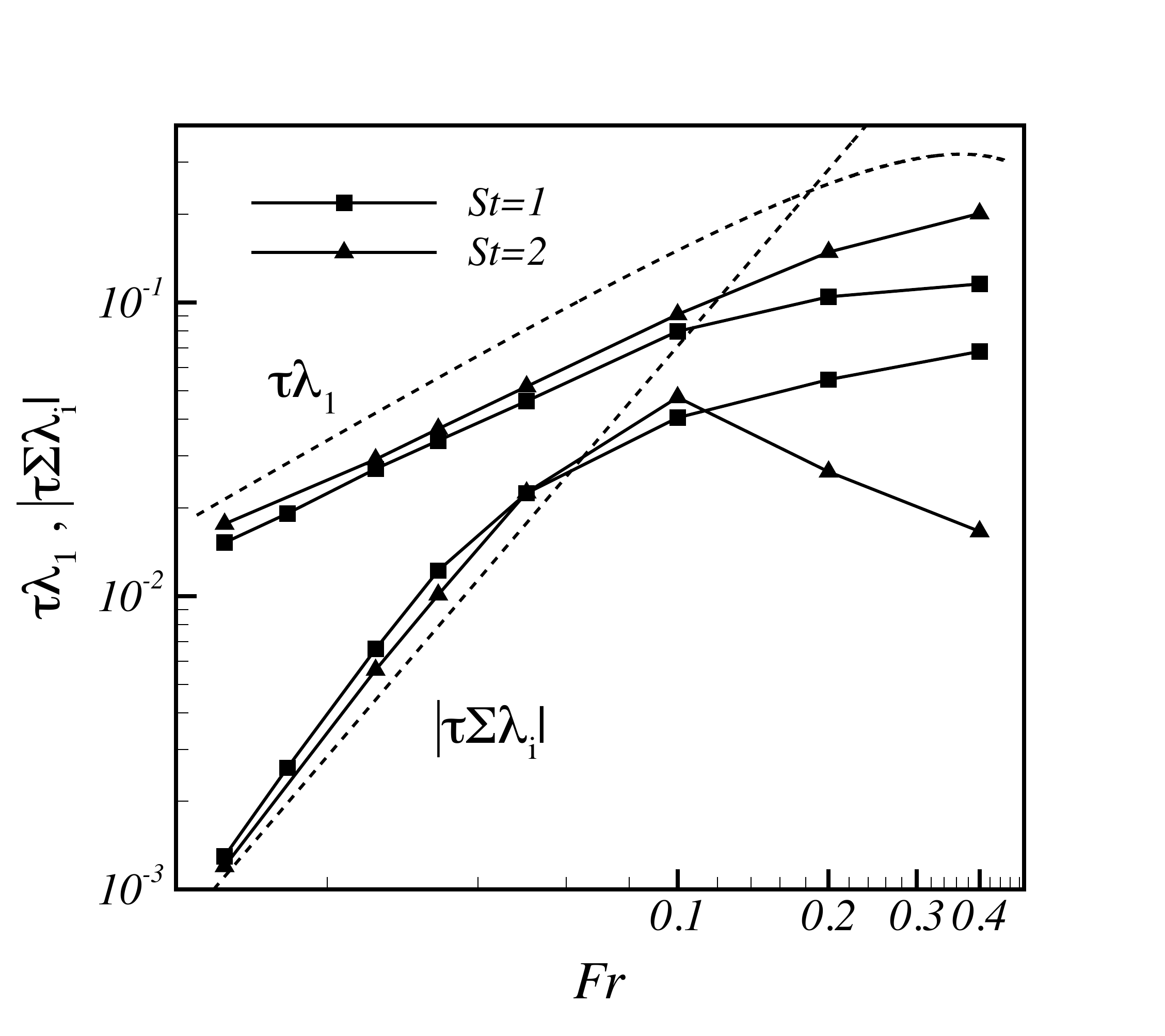}
\caption{Lyapunov exponents obtained from direct simulations of particle-laden isotropic turbulence compared to theoretical predictions (dashed lines, Eqs. \ref{finalf} and \ref{finalsum}). Good agreement between numerical results and theoretical predictions is observed when $Fr \leq 0.03$.} \label{Lyapunov}
\end{figure}

We performed numerical simulation to test theoretical predictions for $\lambda_1 \tau$ (Eq. \ref{finalf}) and $\sum \lambda_i \tau$ (Eq. \ref{finalsum}). Homogeneous isotropic turbulence laden with inertial particles is simulated on a periodic cube. Flow fields is obtained from solving the Navier-Stokes equation using a pseudo-spectral method and the particle motion is computed by taking into account the linear Stokes drag and gravity. To resolve the Kolmogorov length-scale fluid motion, $128^3$ grids are used at the Reynolds number based on the the Taylor scale, $Re_{\lambda}=70$. Information of fluid quantities at the particle position is obtained by the fourth-order Hermite interpolation scheme \cite{CYL, LYC}. Details on numerics can be found in \cite{JYL, CKL, AL1, AL2}. The Lyapunov exponents, $\lambda_1$ and $\sum \lambda_i$, are directly computed by releasing four particles constructing a tetrahedron. The initial distance between particles is set to 1/10,000 of the Kolmogorov length scale and the change of distance between two particles and volume of the tetrahedron is monitored for a period of $45 \tau_{\eta}$ after transient period due to arbitrary initial condition for the particle velocity. 10,000 sets of particles are released in one flow field and data is collected over a total of 23 flow fields. Computed Lyapunov exponents are listed in Table \ref{table1}, in which results for $St=1, Fr=\infty$ (no gravity) are shown for comparison and validation against Bec et al.'s prediction in the absence of gravity \cite{Bec2006}. Theoretical predictions (Eq. \ref{finalf}, \ref{finalsum}) are compared against numerical results in Fig. \ref{Lyapunov}. As predicted, $\sum \lambda_i$ is negative for small $Fr$, and $\lambda_1 \tau$ and $\sum \lambda_i \tau$ do not show $St$-dependency as $Fr \to 0$.

\begin{table*}
\caption{\label{table1} Lyapunov exponents computed from direct numerical simulation of particle-laden turbulence for various values of $St$ and $Fr$. The error range indicates 90 \% confidence level.}
\begin{ruledtabular}
\begin{tabular}{c|c|cccc}
 ~~~~St~~~~ & ~~~~Fr~~~~ &$\lambda_1 \tau$ & $(\sum \lambda_i)\tau$ & $D_{KY}$ & $\lambda_1 \tau/D_{KY}$ \\ \hline 
1 & $\infty$ & 0.1212$\pm 0.0098$ & -0.0809$\pm 0.0456$ & 0.667 & 0.182 \\
1 & 0.4     & 0.1156$\pm 0.0083$   & -0.0681$\pm 0.0244$  & 0.589 & 0.196 \\
1 & 0.2     & 0.1045$\pm 0.0085$   & -0.0546$\pm 0.0327$  & 0.523 & 0.200 \\
1 & 0.1     & 0.0797$\pm 0.0096$   & -0.0405$\pm 0.0073$  & 0.508  & 0.157 \\
1 & 0.05   & 0.0494$\pm 0.0071$   & -0.0240$\pm 0.0131$  & 0.485  & 0.095 \\
1 & 0.033 & 0.0338$\pm 0.0052$   & -0.0122$\pm 0.0069$  & 0.360  & 0.094 \\
1 & 0.025 & 0.0271$\pm 0.0046$   & -0.0066$\pm 0.0036$  & 0.209  & 0.157 \\
1 & 0.0125 & 0.0152$\pm 0.0042$   & -0.0013$\pm 0.0009$  & 0.085  & 0.179 \\ \hline
2 & $\infty$ & 0.2317$\pm 0.0178$ & 0.0629$\pm 0.2298$  & 0.272  & 0.853 \\
2 & 0.4     & 0.2009$\pm 0.0153$   & 0.0166$\pm 0.0752$   & 0.083  & 2.433 \\
2 & 0.2     & 0.1484$\pm 0.0157$   & -0.0265$\pm 0.0554$  & 0.178  & 0.832 \\
2 & 0.1     & 0.0910$\pm 0.0013$   & -0.0475$\pm 0.0352$  & 0.522  & 0.174 \\
2 & 0.05   & 0.0517$\pm 0.0083$   & -0.0225$\pm 0.0136$  & 0.434  & 0.119 \\
2 & 0.033 & 0.0371$\pm 0.0086$   & -0.0101$\pm 0.0065$  & 0.272  & 0.136 \\
2 & 0.025 & 0.0292$\pm 0.0090$   & -0.0056$\pm 0.0028$  & 0.191  & 0.153 \\
2 & 0.0125 & 0.0176$\pm 0.0085$ & -0.0012$\pm 0.0011$  & 0.068  & 0.259 \\

\end{tabular}
\end{ruledtabular}
\end{table*}

We remark that the sum of the Lyapunov exponents is a strictly non-positive quantity which is a consequence of the second law of thermodynamics \cite{FF}. If $\sum\lambda_i>0$ then the total volume of the flow could not be preserved. Thus positive $\sum\lambda_i$ in $St=2$, $Fr=\infty$ case could seem contradictory. This contradiction is lifted by observing that negativity of $\sum\lambda_i$ holds only when there is smooth flow in space. In contrast, there is no flow when $St=2$, $Fr=\infty$. Thus $\sum\lambda_i$ computed as the logarithmic rate of increase of the volume of the tetrahedron built on vectors of three distances between particles does not have implications on other volumes because when there is no smooth flow the tetrahedron's shape is not preserved by the flow. Further, balls of particles do not get deformed into ellipsoids but into volumes of complex shape with possible folding in the same spatial region.  

Returnuing to the case where there is flow, the fact that with probability one volumes of particles decrease exponentially could seem contradictory: in the limit of infinite evolution time the particles would have no volume. This is not a contradiction though but rather this describes the accumulation of particles in space on a time-dependent fractal set with zero three-dimensional volume see details in \cite{Fouxon1}. We now describe the properties of this set. 

\section{Singular distribution of particles in space: fractal dimensions} \label{fractaldimensions}	

In this Section we consider the steady state distribution of particles in space that results in the limit of infinite evolution time. So far we concentrated on the Lyapunov exponents which characterize the temporal phenomena however it is possible to pass from those to the steady state properties. 

The fractal dimension that is based directly on the Lyapunov exponents is the Kaplan-Yorke codimension $D_{KY}$, see \cite{KY}. To write it for particles' flow we observe that $|\lambda_2|\ll \lambda_1$ implies $\lambda_1+\lambda_2>0$. Since $\sum \lambda_i<0$ then the definition \cite{KY} gives for Kaplan-Yorke dimension $D$
\begin{eqnarray}&&
D=2-\frac{\lambda_1+\lambda_2}{\lambda_3}=3-\frac{\sum \lambda_i}{\lambda_3},
\end{eqnarray}
producing for codimension $D_{KY}=3-D$ that $D_{KY} = |\sum \lambda_i|/\lambda_1$ where we use $\lambda_3\approx -\lambda_1$. Using Eqs.~(\ref{finalf}),(\ref{finalsum}) we find
\begin{equation}
D_{KY} =\frac{15\pi \int_0^{\infty}E(k)k dk}{32g}\propto Fr,\label{kydim0}
\end{equation}
Thus $D_{KY}$ is independent of the properties of the particles measured by the Stokes number. Since $Fr\ll 1$ then the Kaplan-Yorke codimension is small $D_{KY}\ll 1$ reflecting the weak compressibility of the flow of particles. Small fractal dimensions signifies that the particles fill the space quite densely. In the limit of strong gravity $Fr\to 0$ the fractal dimension vanishes and the particles fill the space densely (there are particles in the arbitrarily small neighbourhood of arbitrary point in space). The particles in this limit sediment swiftly through space "averaging" the effects of turbulence on their way  

Using the formula (\ref{kydim0}) one can infer the complete spectrum of fractal dimensions described by Eq.~(\ref{spectrumdimensions}). The pair-correlation function's scaling exponent given by Eq.~(\ref{pref}) can be compared with the numerical data of \cite{Becgr}. We find good agreement. The exponent of \cite{Becgr} becomes independent of the Stokes number when $Fr\ll 1$ which agrees with our theoretical prediction. Further the exponent is quite close numerically to $2D_{KY}$ predicted by our theory with $E(k)$ from our numerical simulations.

\section{Cigars or pancakes: the second Lyapunov exponent}

The remaining second Lyapunov exponent describes the logarithmic rate $\lambda_1+\lambda_2$ of growth of areas ($\lambda_1$ describes the growth of line elements, $\lambda_1+\lambda_2$ of area elements and $\lambda_1+\lambda_2+\lambda_3$ of volume). Considered otherwise $\lambda_2$ determines the evolution of the intermediate axis of the ellipsoid into which turbulence deform small volumes of particles. Small ball of particles is deformed into ellipsoid which principal axes $l_i$ obey 
\begin{eqnarray}\!\!\!\!\!\!\!\!\!\!\!\!
&&\lim_{t\to\infty} \frac{\ln l_i(t)}{t}=\lambda_i,
\end{eqnarray}
with $\lambda_1$, $\lambda_3$ defined previously, $\sum \lambda_i$ describing the evolution of the volume $l_1l_2l_3$. The sign of the exponent $\lambda_2$ determines whether the ball evolves into pancake ($\lambda_2>0$) or cigar ($\lambda_2<0$) structure. For passive tracers (fluid particles) $\lambda_2$ is positive \cite{GirPope,Lillo} so pancakes hold. In this Section we demonstrate that for inertial particles $\lambda_2<0$ when $St\gg 1$ and $Fr\ll 1$. Furthermore the magnitude of $\lambda_2$ is quite small $|\lambda_2|/\lambda_1\ll 1$ so the intermediate dimension of the ellipsoid is roughly constant when the growth of $\ln l_1$ at the rate $\lambda_1$ is almost compensated by the decrease of $\ln l_3$ at the rate $\lambda_3\approx -\lambda_1$ providing for parametrically small logarithmic rate of increase of volumes. Similar properties were observed for bubbles when gravity can be neglected \cite{Lillo}. We postpone the interpretation of this similarity to future work.   

Thus volumes of inertial particles are deformed by turbulence into cigars but volumes of fluid particles are deformed into pancakes. This prediction however demands that not only $Fr\ll 1$ but also $St\gg 1$. When $Fr\ll 1$ but $St\sim 1$ our prediction for $\lambda_2$ involves three-point correlation function of turbulent velocity gradients which sign and magnitude are unknown leaving the sign of $\lambda_2$ unknown as well. 

To find $\lambda_2$ we use $\lambda_2=\sum\lambda_i-\lambda_1-\lambda_3$. Using Eqs.~(\ref{rep}), (\ref{rep3}) one finds (cf. \cite{FB})
\begin{eqnarray}\!\!\!\!\!\!\!\!\!\!\!\!
&&\lambda_2=\frac{\sum \lambda_i}{3}-\frac{12}{35}\int_{-\infty}^{t} dt_1dt_2\langle tr s'(t)s'(t_1)s'(t_2)\rangle ,\label{lambda2}
\end{eqnarray}
where to leading order we put $s'$ instead of ${\tilde s}$ in the last term, see the corresponding discussions of $\lambda_1$, $\lambda_3$. 
To leading order in weak compressibility, like in calculations of $\lambda_1$ and $\sum\lambda_i$, we can use the symmetric component $\sigma_l^s=(\sigma_l+\sigma_l^t)/2$ of traceless matrix $\sigma_l$ as $s'$, 
\begin{eqnarray}
&& \!\!\!\!\!\!\!\!\!\!
\lambda_2=\frac{\sum \lambda_i}{3}
-\frac{12}{35}\int_{-\infty}^{t} \langle tr \sigma_l^s(t) \sigma_l^s(t_1) \sigma_l^s(t_2)\rangle dt_1dt_2.
\end{eqnarray}
Using Eq.~(\ref{linear}) we find  
\begin{eqnarray}
&&\!\!\!\!\!\!\!\!\!\!\!\!\lambda_2-\frac{\sum \lambda_i}{3}=-\frac{12}{35}\int_{-\infty}^{t} \frac{dt_1dt_2dt_3}{\tau^3}\int_{-\infty}^{t_1}dt_4\int_{-\infty}^{t_2}dt_5
\nonumber\\&&\!\!\!\!\!\!\!\!\!\!\!\!\langle tr s^s(t_3)s^s(t_4)s^s(t_5)\rangle
\exp\left[\frac{t_3+t_4+t_5-t-t_1-t_2}{\tau}\right],
\end{eqnarray}
where $s^s$ is the symmetric part of $s$.  
Interchanging the orders of integrations over $t_1$ and $t_4$, and $t_2$ and $t_5$ and integrating over $t_1$, $t_2$,
\begin{eqnarray}
&&\!\!\!\!\!\!\!\!\!\!\!\!\lambda_2-\frac{\sum \lambda_i}{3}=-\frac{12}{35}\int_{-\infty}^{t} \frac{dt_3dt_4dt_5}{\tau}\langle tr s^s(t_3)s^s(t_4)s^s(t_5)\rangle
\nonumber\\&&\!\!\!\!\!\!\!\!\!\!\!\!
\biggl(\exp\left[\frac{t_3+t_4+t_5-3t}{\tau}\right]-\exp\left[\frac{t_3+t_4-2t}{\tau}\right]
\nonumber\\&&
-\exp\left[\frac{t_3+t_5-2t}{\tau}\right]+\exp\left[\frac{t_3-t}{\tau}\right]\biggr).\nonumber
\end{eqnarray}
Using that the correlation function $\langle tr s^s(t_3)s^s(t_4)s^s(t_5)\rangle$ vanishes rapidly when $|t_4-t_3|$ or $|t_5-t_3|$ become much greater than $\tau_g\ll \tau$ we find 
\begin{eqnarray}
&&\!\!\!\!\!\!\!\!\!\!\!\!\lambda_2-\frac{\sum \lambda_i}{3}=-\frac{12}{35}\int_{-\infty}^{t} \frac{dt_3dt_4dt_5}{\tau}\langle tr s^s(t_3)s^s(t_4)s^s(t_5)\rangle
\nonumber\\&&\!\!\!\!\!\!\!\!\!\!\!\!
\biggl(\exp\left[\frac{3t_3-3t}{\tau}\right]-2\exp\left[\frac{2t_3-2t}{\tau}\right]
+\exp\left[\frac{t_3-t}{\tau}\right]\biggr).\nonumber
\end{eqnarray}
We observe that except for $t_3$ in the $\tau_g-$vicinity of $t$ the integral $\int^t dt_4dt_5\langle tr s^s(t_3)s^s(t_4)s^s(t_5)\rangle$ is a constant which we designate by $c_0$ where 
\begin{eqnarray}
&&c_0=\int_{-\infty}^{\infty} dt_4dt_5\langle tr s^s(t_3)s^s(t_4)s^s(t_5)\rangle
\end{eqnarray}
Neglecting that vicinity in comparison with the much larger size $\tau$ of the effective domain of integration over $t_3$ we find  
\begin{eqnarray}
&&\!\!\!\!\!\!\!\!\!\!\!\!\lambda_2-\frac{\sum \lambda_i}{3}\approx -\frac{12c_0}{35}\int_{-\infty}^{t} \frac{dt_3}{\tau}\biggl(\exp\left[\frac{3t_3-3t}{\tau}\right]\nonumber\\&&\!\!\!\!\!\!\!\!\!\!\!\!
-2\exp\left[\frac{2t_3-2t}{\tau}\right]
+\exp\left[\frac{t_3-t}{\tau}\right]\biggr)=-\frac{4c_0}{35}.\nonumber
\end{eqnarray}
Finally rewriting $c_0$ in terms of the instantaneous correlation functions of turbulence like in calculations of $\lambda_1$, $\sum\lambda_i$ we find 
\begin{eqnarray}
&&\!\!\!\!\!\!\!\!\!\!\!\!\lambda_2\tau\!=\!-\!
\int_{-\infty}^{\infty}\!\! \frac{4dx' dx''}{35 g^2\tau}\langle tr (\nabla \bm u)_s(0) (\nabla \bm u)_s(x') (\nabla \bm u)_s(x'')\rangle \nonumber\\&&\!\!\!\!\!\!\!\!\!\!\!\!
+\frac{\sum \lambda_i\tau}{3}, \label{sec}
\end{eqnarray}
where $(\nabla \bm u)_{s, ik}=[\nabla_iu_k+\nabla_k u_i]/2$ is the strain matrix of the turbulent flow. We observe that the RHS in the first line is of order of $Fr^2/St$. Since $\sum \lambda_i\tau\sim Fr^2$ we conclude that 
\begin{eqnarray}
&&\!\!\!\!\!\!\!\!\!\!\!\!\lambda_2\tau=\frac{\sum \lambda_i\tau}{3}=-\frac{[\pi\int E(k)kdk]^2}{32g^2}\propto Fr^2
,\ \ St\gg 1. \label{final2}
\end{eqnarray}
Clearly $\lambda_2/\lambda_1\propto Fr\ll 1$ where we used $\lambda_1\tau\propto Fr$. 

Thus when $St\gg 1$ the second Lyapunov exponent has the same degree of universality as the rest of $\lambda_i$ being determined by $\int E(k) kdk$. 
However when $St\sim 1$ both terms in Eq.~(\ref{sec}) are of the same order $Fr^2$ and have to be kept. Thus in the latter case the answer for $\lambda_2$ is not describable completely by $\int E(k) kdk$ in contrast to $\lambda_1$ and $\sum\lambda_i$. Further though $\lambda_2/\lambda_1\propto Fr$ continues to hold when $St\sim 1$ the sign of $\lambda_2$ can no longer be obtained from Eq.~(\ref{sec}) because it seems not possible to fix the sign of RHS in the first line.

\section{$St\gtrsim 1$ and $Fr\gtrsim 1$ is the boundary for formation of the fractal}

We demonstrated that when $St\gtrsim 1$ and $Fr\ll 1$ the particles concentrate on the fractal set. We now demonstrate that when the Froude number is increased at fixed Stokes number $St\gtrsim 1$ then there is a transition at $Fr_{cr}\sim 1$ where particles become space-filling. Note that only weak dependence of $Fr_{cr}$ on $St$ is expected. 

We observe that when the Froude number is increased from $Fr\ll 1$ our previous considerations imply that blow ups of the gradients of $\bm v$ governed by Eq.~(\ref{eff_vel_cond}) happen more often. These blow ups are nothing but the sling events where the particles thrown by the flow start to move inertially colliding with other point particles, see Section \ref{theory}. This is because the blow ups correspond to the evolution of $\sigma$ given by ${\dot \sigma}+\sigma^2=0$ that comes from differentiating the equation of motion of free particles $\partial_t\bm v+(\bm v\cdot\nabla)\bm v=0$, cf. \cite{FFS1}. Thus at $Fr\sim 1$ the events of particles detaching from the local flow and moving ballistically become typical. These events destroy the fractal structure producing particles' distribution that fills the whole space. 

\begin{figure}
\includegraphics[width=8.0 cm,clip=]{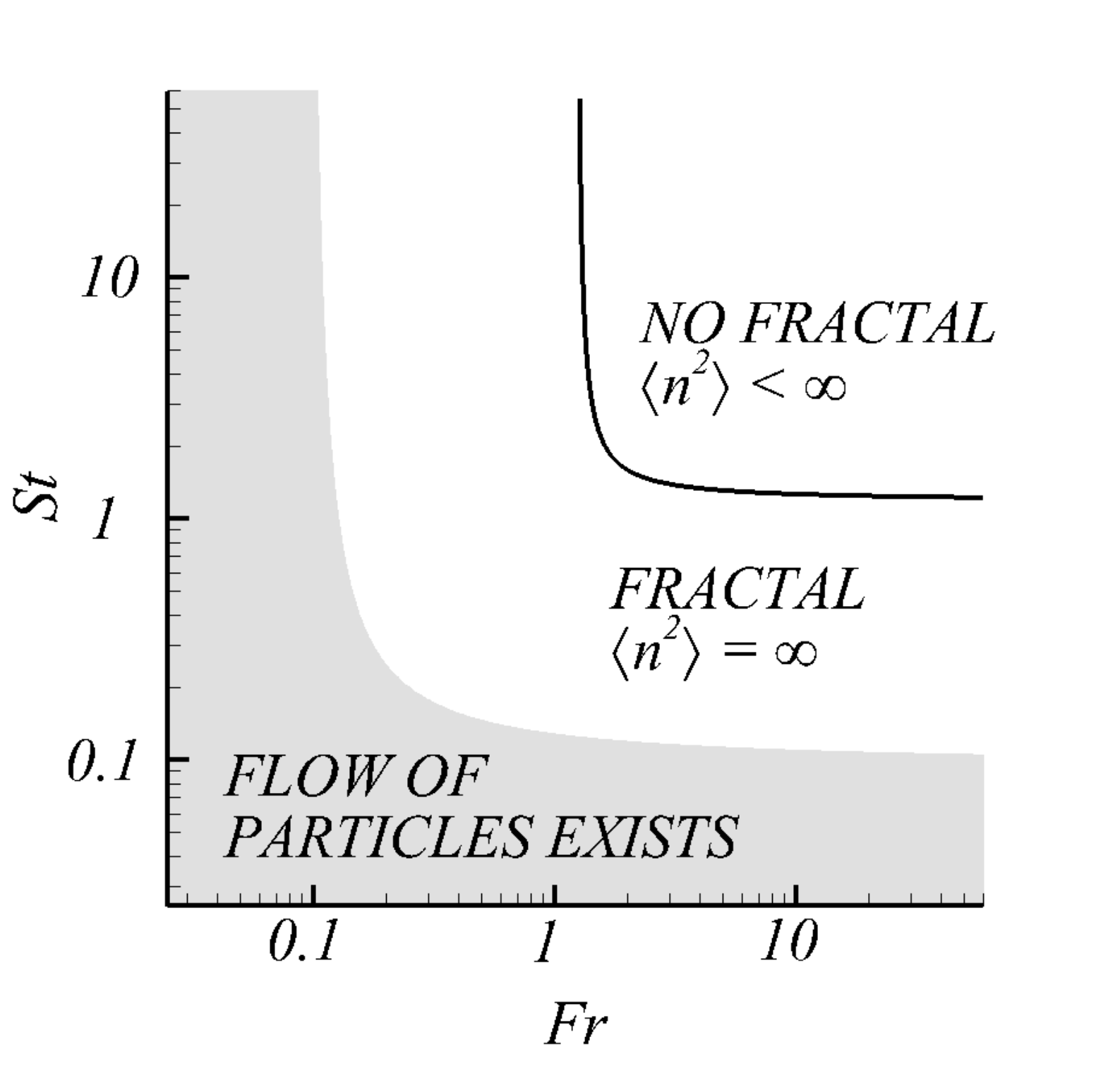}
\caption{In the region $\min[Fr, St]\ll 1$ one can introduce the flow of particles implying that fractality of the spatial distribution. The line separating the region of the flow from the region where velocity is significantly multi-valued is not sharp. In contrast the transition from the case of infinite $\langle n^2\rangle$ to finite $\langle n^2\rangle$ defines a sharp transition line in the plane of dimensionless parameters of the problem.} \label{fractal}
\end{figure}

The transition occurs similarly to the transition at $Fr\to\infty$ when gravity can be disregarded. Then at $St\ll 1$ one can introduce the flow of the particles as described by Eq.~(\ref{eff_vel}) with $v_g$ set to zero. However when $St\sim 1$ the probability of blow ups is not negligible so that the flow can no longer be introduced. This however does not say that fractality is destroyed - simply the role of ballistic events grows. The transition to the non-fractal structure with finite value of $\langle n^2\rangle$ occurs at $St\sim 1$, see \cite{BecCenciniHillerbranddelta}. Similar scenario holds due to increase of $Fr$. The schematic plot of the phase diagram in the plane of the dimensionless parameters $(Fr, St)$ can be seen in Figure (\ref{fractal}), cf. \cite{FHa}.

\section{The flow at $St\ll 1$ in the presence of gravity}\label{smallSt}

So far we concentrated mostly on the case of $St\gtrsim 1$ where the role of gravity is crucial - it causes the irregular, ballistic-like spatial motion of particles at $St\gg 1$ that consists of many streams intersecting at the same point to become smooth and single-valued. In the case of $St\ll 1$ the flow holds already without gravity so the role of the latter is not that decisive. Gravity does not break the condition for the existence of the particles' flow at $St\ll 1$: since $\min[\tau, \tau_\eta, \tau_{d}]\leq \tau$ then at $St\ll 1$ we have $\tau \min[\tau, \tau_\eta, \tau_{d}]\leq \tau^2\ll \tau_\eta^2$ so that the condition (\ref{condition}) holds irrespective of the Froude number. Correspondingly the universal description described in Sec. \ref{theory} holds at $St\ll 1$ irrespective of the Froude number. However the flow itself can be changed by gravity significantly. In this Section we consider how gravity changes the Lyapunov spectrum and $D_{KY}$. Some results of this Section were derived previously in \cite{FFS,FHa}.

We demonstrated in Secs. \ref{spectrum}-\ref{third} that when $St\gg Fr$ we have $\lambda_3\approx -\lambda_1$ with $\lambda_1$ given by Eq.~(\ref{finalf}). Further, gravity is negligible when $St\ll Fr$ (which implies $St\ll 1$ since $Fr\ll 1$) so that $\lambda_i$ are close to Lyapunov exponents of fluid particles and $D_{KY}$ is given by Eq.~(\ref{kydim01}). Thus it remains to consider $D_{KY}$ and $\lambda_2$ when $Fr\ll St\ll 1$. 

We start with the case of $St\gg Fr$ but $St^2\ll Fr$ where $\tau\ll\tau_g\ll \tau_{\eta}$. In this case one can use Eqs.~(\ref{eff_vel}) and (\ref{kydim1}). The temporal correlations of the gradients are due to the gravitational drift so that 
\begin{eqnarray}&&
\int \left\langle \phi[0, 0] \phi[t, \bm x_w(t, 0)] \right\rangle dt \approx \int F[\bm g \tau t] dt=\frac{\int F(x) dx}{g\tau},\nonumber
\end{eqnarray}
where $F(x) \equiv \left\langle \phi(0) \phi(\bm x) \right\rangle$ is the single-time correlation function of $\phi$. Since $\phi =- \nabla^2 p$, we see that $F(x) =\nabla^4 \left\langle p(0)p(\bm x)\right\rangle$. Introducing the spectrum of fluctuations of pressure $E_p(\bm k)$ by 
\begin{eqnarray}&&
\langle p(0)p(\bm x)\rangle=\int E_p(k)\exp[i\bm k\cdot\bm x]\frac{d\bm k}{4\pi k^2},
\end{eqnarray}
so that $\langle p^2\rangle=\int_0^{\infty} E_p(k)dk$ we have 
\begin{eqnarray}&&\!\!\!\!\!\!\!\!\!
F\!=\!\int\! k^2 E_p(k)\exp[i\bm k\cdot\bm x]\frac{d\bm k}{4\pi}\!=\!\int_0^{\infty}\! \frac{k^3 E_p(k)\sin(kx)dk}{x},\nonumber
\end{eqnarray}
which gives 
\begin{eqnarray}&&\!\!\!\!\!\!\!\!\!
\int F(x)dx=\pi \int_0^{\infty} k^3 E_p(k)dk.
\end{eqnarray}
Using Eqs.~(\ref{kydim1}),(\ref{finalf}) with $|\lambda_3^w|=\lambda_1$ we find
\begin{eqnarray}&&\!\!\!\!\!\!\!\!\!\!\!\!\!\!\!
D_{KY}\! = \!\frac{5\tau^2 \int_0^{\infty}\! k^3 E_p(k)dk}{2\int_0^{\infty} E(k)kdk}\propto St^2,\ \ St^2\ll Fr\ll St. \label{dimsn} 
\end{eqnarray}
Thus though the gravity changes $D_{KY}$ it disappears from the final answer. We observe that at $St\ll 1$ both when $St\ll Fr$ (when $D_{KY}$ is given by Eq.~\ref{kydim01}) and $St^2\ll Fr\ll St$ the Kaplan-Yorke dimension is proportional to $St^2$ however the proportionality coefficient is different. In fact, due to intermittency the proportionality coefficient is a power of Reynolds number so the difference is parametric in $Re\to\infty$ limit, cf. \cite{FP1,FP2}. When $Re$ is moderate the proportionality coefficients are likely to be comparable. 

We recall that flow gradients are short correlated when $St\gg Fr$, see Sec. \ref{spectrum}. 
Thus to find the second Lyapunov exponent we can use Eq.~(\ref{lambda2}) where at $St^2\ll Fr\ll St$ we can use the strain matrix $\nabla \bm u_s$ instead of $s'$. Since the correlation time of $\nabla \bm u_s$ in the particle's frame is $\tau_g$ then the second term on the RHS of Eq.~(\ref{lambda2}) is of order 
$\tau_g^2/\tau_{\eta}^3\sim \tau^{-1}Fr^2/St$ which ratio to $|\sum\lambda_i|=D_{KY}\lambda_1\sim \tau^{-1}St^2Fr$ is of order $Fr/St^3\gg 1$ (it is the product of two large numbers $Fr/St^2$ and $1/St$). Thus $\lambda_2\tau$ is of order $Fr^2/St$ and dominated by the second term on the RHS of Eq.~(\ref{lambda2})
\begin{eqnarray}\!\!\!\!\!\!\!\!\!\!\!\!
&&\lambda_2\!=\!-\!
\int_{-\infty}^{\infty}\!\! \frac{12 dx' dx''}{35 g^2\tau^2}\langle tr (\nabla \bm u)_s(0) (\nabla \bm u)_s(x') (\nabla \bm u)_s(x'')\rangle.\nonumber
\end{eqnarray}
We observe that $\lambda_2/\lambda_1$ which smallness follows from smallness of correlation time of $\sigma$ obeys $\lambda_2/\lambda_1\sim Fr/St\ll 1$. Clearly this is also the order of correction to $\lambda_1$ given by Eq.~(\ref{finalf}).

Finally we consider the remaining asymptotic region $\tau_g\ll \tau\ll \tau_{\eta}$ (which is $Fr\ll St^2$). This is the case where the smallest time-scale of fluctuations of velocity in the particle's frame $\tau_g$ is smaller than the time $\tau$ in Eq.~(\ref{vel_point}) so that the particle's velocity dependence on  $\bm u(t, \bm x)$ is non-local in time. Thus no explicit expression for the particles' flow $\bm v(t, \bm x)$ is available. In particular, equation \eqref{eff_vel} breaks down. This situation is similar to the case of $St\gtrsim 1$, $Fr\ll 1$ that we studied previously. In fact, considering the calculation of $\sum\lambda_i$ in Sec. \ref{sumsection} one finds that it holds completely when $Fr\ll St^2$ as well. The Gaussianity of $\sigma$ follows from Eq.~(\ref{linear}) by $\tau_g\ll \tau$. The smallness of the correlation time $\tau$ follows from the condition $|\sigma|\tau\ll 1$ of the flow existence. We conclude that
\begin{equation}
D_{KY} =\frac{15\pi \int_0^{\infty}E(k)k dk}{32g}\propto Fr,\ \ St^2\gg Fr,
\end{equation}
where the applicability condition $St^2\gg Fr$ includes the previously found range $Sr\gtrsim 1$, $Fr\ll 1$. 

To find $\lambda_2$ in the range $St^2\gg Fr$ we take again Eq.~(\ref{lambda2}) as the starting point. Using $\tau_g\ll \tau$ one finds that the calculations of Sec. hold so that $\lambda_2$ is given by Eq.~(\ref{sec}). Since the first term in the latter formula is of order $\tau^{-1}Fr^2/St$ and the second is of order $\tau^{-1}Fr^2$ then $\lambda_2$ obeys 
\begin{eqnarray}\!\!\!\!\!\!\!\!\!\!\!\!
&&\lambda_2\!=\!-\!\int_{-\infty}^{\infty}\!\! \frac{4 dx' dx''}{35 g^2\tau^2}\langle tr (\nabla \bm u)_s(0) (\nabla \bm u)_s(x') (\nabla \bm u)_s(x'')\rangle.\nonumber
\end{eqnarray}
when $St^2\gg Fr$ but $St\ll 1$. It is not possible to fix the sign of $\lambda_2$. 

\section{Cross-correlation of positions of particles and vortices}

The term ``preferential concentration" would signify that particles prefer to concentrate in particular regions of the flow. However the quantities considered so far described the spatial distribution of the particles without the discussion of how it is correlated with the spatial distribution of the vortices. The latter correlation can be obtained in the case $\tau\ll \min[\tau_{\eta}, \tau_g]$ where Eq.~(\ref{eff_vel}) holds. This is the case of $St\ll 1$, $Fr\gg St^2$. It is possible to use then the consideration of \cite{Fouxon1} to demonstrate that the steady state density $n_s$ obeys 
\begin{eqnarray}&&\!\!\!\!\!\!\!\!\!\!\!\!\!\!
\int \left[w'^2(\bm x)-s_u^2(\bm x)\right]n_s(\bm x)d\bm x=\frac{\sum \lambda_i}{\tau}, 
\end{eqnarray}
where $w'$ is the vorticity of turbulence, $s_u$ is the strain of turbulence and $\sum \lambda_i<0$, see \cite{Fouxon1}. This equation is the statement that $\lim_{t\to\infty}\langle \nabla\cdot \bm v[t, \bm X(t, 0)]\rangle=\sum\lambda_i$. It says that particles sample the flow so that they are predominantly in the regions dominated by the strain: if $\bm x_i$ are particles' positions then  
\begin{eqnarray}&&\!\!\!\!\!\!\!\!\!\!\!\!\!\!
\lim_{N\to\infty}\frac{1}{N}\sum_{i=1}^N\left[w'^2(\bm x_i)-s_u^2(\bm x_i)\right]=\frac{\sum \lambda_i}{\tau}, \label{sumlambda0}
\end{eqnarray}
where the number of particles $N$ is to be large. Using the previously found results for $\sum \lambda_i$ we have
\begin{eqnarray}&&\!\!\!\!\!\!\!\!\!\!\!\!\!\!\!\!\!\!\!
\lim_{N\to\infty}\sum_{i=1}^N\frac{\tau_{\eta}^{2}\left[s_u^2(\bm x_i)\!-\!w'^2(\bm x_i)\right]}{N}\! \sim \!St,\ \  \!\!St\!\!\lesssim \!Fr,\label{sumlambda1}\\&&\!\!\!\!\!\!\!\!\!\!\!\!\!\!\!\!\!\!\!
\lim_{N\to\infty}\sum_{i=1}^N\frac{\tau_{\eta}^{2}\left[s_u^2(\bm x_i)\!-\!w'^2(\bm x_i)\right]}{N} \!\sim \!Fr,\ \ \!\! St^2 \!\!\ll \!Fr\!\!\ll\!\! St.\label{sumlambda2}
\end{eqnarray}
providing the measure of anti-correlations between the locations of the particles and the vorticity divided by its typical value $1/\tau_{\eta}$.

\begin{figure}
\includegraphics[width=8.0 cm,clip=]{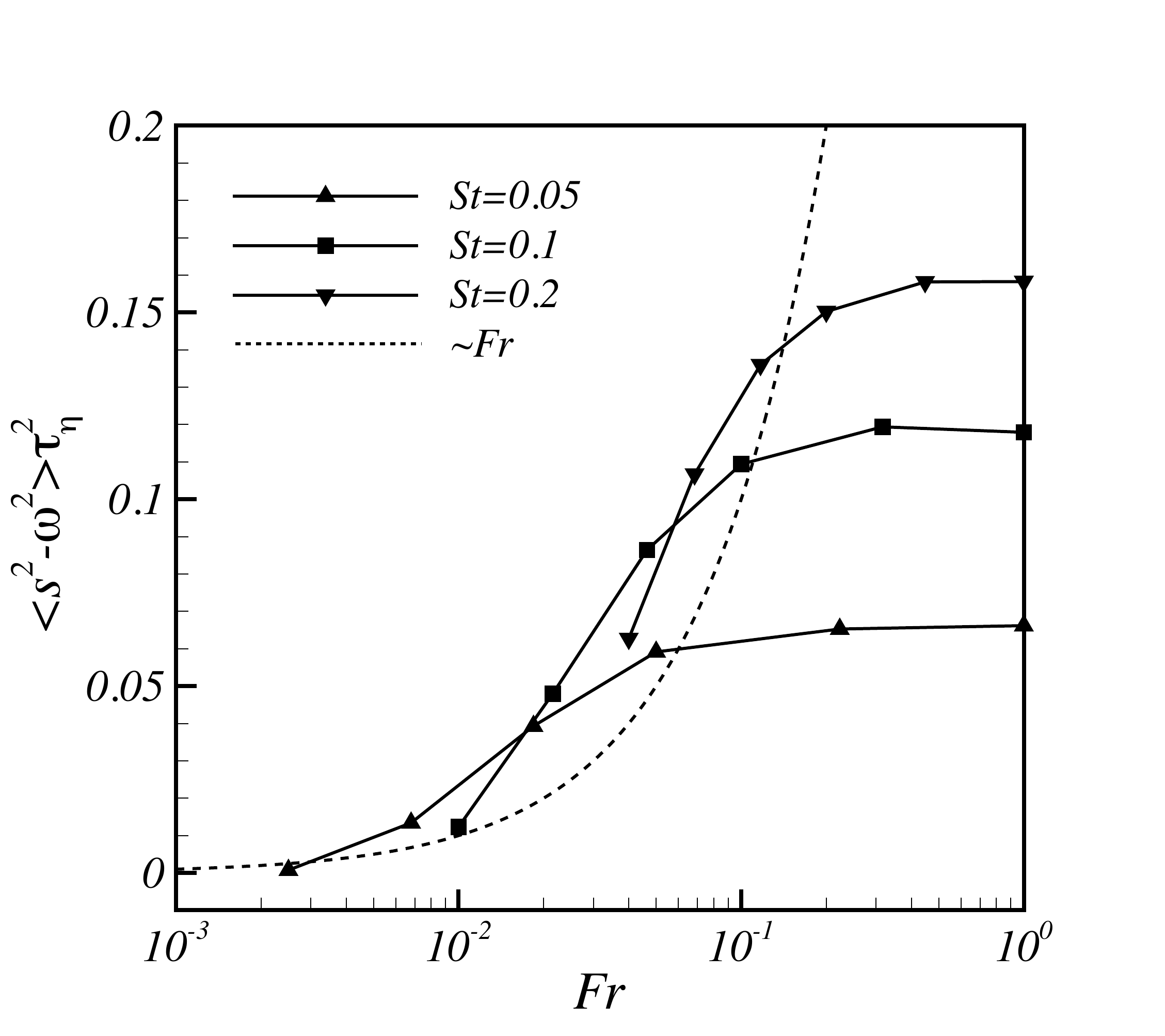}
\caption{The sum of the Lyapunov exponents numerically obtained for $St=0.05, 0.1$ and 0.2. When $Fr$ becomes larger than $St$, the effect of gravity becomes negligible and curves are predicted to tend to a constant proportional to $St$ provided $St\ll 1$. The slight deviations from this prediction at $St=0.2$ are explained by violation of the condition $St\ll 1$. The range of parameters $St^2\ll Fr\ll St$, where theoretical predictions exist, demands very small $St$ and holds roughly for $St=0.05$ where numerical results agree with theoretical predictions. } \label{laplaceP}
\end{figure}

Here, we compare theoretical estimation (Eqs. \ref{sumlambda1}, \ref{sumlambda2}) with numerical results obtained from direct simulation of particle-laden turbulence. For three values of Stokes number smaller than 1, $St=0.05, 0.1, 0.2$, the normalized sum of the Lyapunov exponents computed using Eq. \ref{sumlambda0} in terms of the Froude number are illustrated in Fig. \ref{laplaceP}. All other parameters in turbulence simulation are the same as before (Section \ref{sumsection}). For the range of $Fr \gtrsim St$, the normalized sum of Lyapunov exponent approaches the value of order 1 with some scatters. When $St^2 \ll Fr \ll St$, the linear relation roughly holds for all Stokes numbers, confirming the theory.

We now consider the cross-correlation in the regions $St\gtrsim 1$, $Fr\ll 1$ or $St\ll 1$, $Fr\ll St^2$. In both these regions one can introduce the flow of particles but cannot write it explicitly. In both these regions the correlation time of $s(t)$ is $\tau_g$ and it is much less than $\tau$. 
Noting that $tr s^2=s_u^2-w'^2$ we have 
\begin{eqnarray}&&
\int \left[w'^2(\bm x)-s_u^2(\bm x)\right]n_s(\bm x)d\bm x=-\langle n_s  tr s^2\rangle
\nonumber\\&&
=-\left\langle tr s^2(0)\exp\left[-\int_{-\infty}^0 tr\sigma(t) dt
\right]\right\rangle,
\end{eqnarray}
where we used the representation of the steady state density introduced in \cite{Fouxon1}. To lowest order in weak compressibility we find 
\begin{eqnarray}&&\!\!\!\!\!\!\!\!\!\!\!\!
-\langle n_s  tr s^2\rangle=\int_{-\infty}^0\langle tr s^2(0) tr\sigma(t)\rangle dt
\nonumber\\&&\!\!\!\!\!\!\!\!\!\!\!\!
=\int_{-\infty}^0 dt\int_{-\infty}^t dt'\langle tr s^2(0)tr \sigma_l^2(t')\rangle \exp \left( \frac{t'-t}{\tau} \right),
\end{eqnarray}
where we used Eq.~(\ref{trace}). Using the definition of $\sigma_l$ in Eq.~(\ref{linear}) we have
\begin{eqnarray}&&
-\langle n_s  tr s^2\rangle=
-\frac{2}{\tau^2}\int_{-\infty}^0 dt\int_{-\infty}^t dt'\int_{-\infty}^{t'} dt_1\int_{-\infty}^{t_1}dt_2 
\nonumber\\&&\exp \left( \frac{t_1+t_2-t'-t}{\tau} \right)\langle tr s^2(0)tr s(t_1)s(t_2)\rangle,
\end{eqnarray}
where we used the symmetry of the integrand in $t_1$, $t_2$. To lowest order in weak compressibility one can use in the above equation the trajectories of the incompressible component of the flow. 
We observe that since $\langle s_{ik}\rangle=0$ and $\langle tr s^2(0)\rangle=0$ then $\langle tr s^2(0)tr s(t_1)s(t_2)\rangle$ is
non-vanishing only of $|t_1|\lesssim \tau_g$, $|t_2|\lesssim \tau_g$. It is then readily seen that $t$, $t'$ in the domain of integration obey 
$|t|\lesssim \tau_g$, $|t'|\lesssim \tau_g$ too. Thus the exponent can be set to one in the integrand. Interchanging the order of integrations 
\begin{eqnarray}&&
-\langle n_s  tr s^2\rangle=
-\frac{2}{\tau^2}\int_{-\infty}^0 dt_2\int_{t_2}^{0} dt_1\int_{t_1}^{0} dt'\int_{t'}^{0}dt 
\nonumber\\&&\langle tr s^2(0)tr s(t_1)s(t_2)\rangle
\nonumber\\&&
=-\frac{1}{\tau^2}\int_{-\infty}^0 dt_2\int_{t_2}^{0} dt_1 t_1^2
\langle tr s^2(0)tr s(t_1)s(t_2)\rangle.
\end{eqnarray}
Finally we rewrite the answer in terms of the single-time correlation functions similarly to previous study
\begin{eqnarray}&&\!\!\!\!\!\!\!
\tau_{\eta}^2\int \left[w'^2(\bm x)-s'^2(\bm x)\right]n_s(\bm x)d\bm x=-\frac{\tau_{\eta}^2}{\tau^2}\int_{-\infty}^0 dt_2\int_{t_2}^{0} dt_1 
\nonumber\\&&\!\!\!\!\!\!\!
t_1^2 \langle \nabla_k u_i(0)\nabla_i u_k(0)\nabla_p u_r(\bm g\tau t_1)\nabla_r u_p(\bm g \tau t_2)\rangle.
\end{eqnarray}
This provides the cross-correlations in terms of the properties of turbulence. 
The RHS can be estimated as $\tau_g^4/\tau^2\tau_{\eta}^2=Fr^4/St^6$.
It is quite likely that the RHS is negative so that the anti-correlations between the location of the particles and the vortices persist into the region where the flow exists but cannot be written explicitly. However it seems that the only way to check the RHS's sign it to compute the corresponding fourth-order correlation function of the turbulent velocity gradients which is beyond the scope of this work. 

\section{Role of the Reynolds number}

The problem of distribution of inertial particles in turbulence is determined by three dimensionless parameters. So far we considered the dependence on two of those parameters - the Stokes and the Froude numbers, ignoring the dependence on the Reynolds number. This is tantamount to disregarding the intermittency of turbulence presuming that the statistics of the gradients of the turbulent flow is universal when gradients and inverse time-scales are measured in the units of $\sqrt{\epsilon/\nu}$. In real flows that statistics is not universal but rather depends on the Reynolds number due to intermittency. Intermittency increases the answers in Eqs.~(\ref{kydim01}), (\ref{kydim1}) that contain quartic (effectively triple due to time integration) moments of velocity gradients by a power of $Re$ see \cite{FFS1,FP2}. In contrast the answers provided in cases $Fr\ll St^2\ll 1$ and $St\gtrsim 1$ but $Fr\ll 1$ involve only quadratic moments and will depend on $Re$ very weakly (the dependence is still there due to fluctuations of the viscous range cf. \cite{Frisch}). Similarly the answers for the cross-correlations of the locations of the particles and the vortices involve higher order moments and thus a power of $Re$.

\section{Conclusions and discussion}

In this paper we solved the problem of spatial distribution of inertial particles that settle gravitationally in turbulence with small Froude number. 
We stress that smallness of Froude number is the property of turbulence and not of particles: our results hold for arbitrary particles dragged linearly by the flow with typical acceleration of fluid particles $\epsilon^{3/4}/\nu^{1/4}$ much smaller than the gravitational acceleration $g$. The Froude number is the accelerations' ratio $Fr=\epsilon^{3/4}/[g\nu^{1/4}]$. Thus particles with different size and densities (Stokes numbers) can be uniformly described by our results.

Before passing to the description of the results we observe that the considered limit is of direct practical value: the situation of turbulence with small Froude numbers is typical for air turbulence in warm clouds. The latter increases the collision rate of water droplets thus increasing the rate of rain formation non-negligibly see e. g. \cite{review} and references therein. Using the viscosity of air of $1.8*10^{-4}$ $cm^2/s$ one has $Fr\sim \epsilon^{3/4}10^{-3}$ where $\epsilon$ is measured in $cm^2/s^3$. Since typically $\epsilon$ ranges from $10 cm^2/s^3$ for stratocumulus and $100 cm^2/s^3$ for small cumulus clouds \cite{review} then this number is quite small. Thus our results hold for water droplets in clouds provided their size is not too large so that the particles' driving by the flow can be described by the linear drag. This holds true for droplets with size smaller than $30 \mu m$. Since turbulence is relevant for droplets in the range from $15$ to $30$ microns then our theory holds in the range of interest (for sizes larger than $30 \mu m$ droplet growth proceeds rather by gravitational collision-coalescence). We demonstrated by numerical simulations that the theoretical predictions hold well when $Fr\leq 0.03$. We hope thus that the results described below like the one for preferential concentration can be readily incorporated into droplets' collision kernels used in numerical simulations to become a tool in rain formation prediction. 

We concentrate the results that hold for all particles driven by small Froude number turbulence with the driving force describable by the linear drag law. In the steady state the particles distribute over fractal set in space which changes in time keeping its spatial statistics intact. The statistics of this set is log-normal being described by one number only - the Kaplan-Yorke dimension $D_{KY}$ which is much smaller than one \cite{Fouxon1}. In terms of this number the correlation functions of the spatial concentration of the particles $n$ obey ($\langle n\rangle=1$)
\begin{eqnarray}&&\!\!\!\!\!\!\!\!\!\!\!\!\!\!
\left\langle n(\bm x_1) n(\bm x_2) \ldots n(\bm x_N) \right\rangle \! =\prod_{i>k}\left( \frac{\eta}{|\bm x_i-\bm x_k|} \right)^{2D_{KY}}.\label{multipoint}
\end{eqnarray}
where $|\bm x_i-\bm x_k|\ll \eta$. Preferential concentration is conventionally described by the pair-correlation function 
\begin{eqnarray}&&
\left\langle n_s(0) n_s(\bm x) \right\rangle = \left( \frac{\eta}{x} \right)^{2D_{KY}},
\ \
x \ll \eta.\label{pair}
\end{eqnarray}
that describes the enhancement of the probability of two particles to be at close distance $r$ due to inertia. Since $D_{KY} \ll 1$ then $\left\langle n_s(0) n_s(\bm x) \right\rangle\approx \left\langle n_s(0) \right\rangle \left\langle n_s(\bm x) \right\rangle =1$ unless $x$ is significantly smaller than $\eta$. Thus the density fluctuates significantly only at scales much smaller than $\eta$ and equation \eqref{multipoint} completely determines the statistics in the region where the fluctuations are significant. 

The correlation functions diverge at fusion of the points $|\bm x_i-\bm x_k|\to 0$. This divergence is the consequence of the singularity of the spatial distribution of the particles. When one tries to define concentration $n(\bm x)$ by counting the number of particles $m_l(\bm x)$ in the ball of small radius $l$ centered at $\bm x$, dividing by the volume $4\pi l^3/3$  and taking $l\to 0$ limit one finds that the limit is either zero or infinity providing no useful information. In contrast the logarithm of the coarse-grained density $n_l(\bm x)=3m_l(\bm x)/4\pi l^3$ divided by $\ln (l/\eta)$ has finite limit,  
\begin{equation}
\lim_{l\to 0}\frac{\ln n_l(t, \bm x)}{\ln(l/\eta)}=D_{KY},\label{limit}
\end{equation}
which is the same for all $\bm x$ except for those which have zero spatial volume. (This limit is direct consequence of formulas derived in \cite{Fouxon1} see \cite{lectures} for details.) The latter $\bm x$ are non-negligible because they include the fractal containing the particles.
Indeed since $D_{KY}>0$ then $n_s(\bm x)=0$ for $\bm x$ for which the limit (\ref{limit}) holds. However for arbitrarily small but finite $l$ one has strong fluctuations of $n_l(\bm x)$ described by the log-normal distribution \cite{Fouxon1} 
\begin{equation}
\langle  n_l^{\rho}\rangle= \left( \frac{\eta}{l} \right)^{D_{KY}\rho(\rho-1)},\label{moments}
\end{equation}
where $\rho=2$ case reproduces Eq.~(\ref{pair}). Since the zero total volume points $\bm x$ for which Eq.~(\ref{limit}) does not hold can be neglected in the spatial averages $\langle  n_l^{\rho}\rangle$ of finite field $n_l(\bm x)$ then we $n_l(\bm x)$ is a strongly intermittent field whose moments are determined by rare strong fluctuations. This field can be described by multi-fractal model that in contrast to similar model for turbulence \cite{Frisch} holds for inertial particles rigorously \cite{lectures}. Using the result (\ref{moments}) one can readily determine the spectrum of fractal dimensions $D(\alpha)$ defined by \cite{HP, BGH}
\begin{equation}
D(\alpha) \equiv \lim_{l \to 0} \ln \frac{\left\langle m_l^{\alpha-1} n_{s} \right\rangle}{(\alpha - 1) \ln l}= 3 - D_{KY} \alpha \ ,
\end{equation}
These results hold for all particles in turbulence with small Froude number. They provide complete description of the spatial statistics of particles in terms of one unknown number $D_{KY}$. 

Clearly the results signal the presence of underlying universal description. This description is the one of smooth flow in space which has small compressibility and finite correlation time. In this work we demonstrated that the motion of particles in turbulence with small Froude number can be described by this kind of flow. 

To complete the description it remains to find $D_{KY}$ which at fixed $Fr\ll 1$ is generally determined by both the statistics of the carrying turbulent flow and the properties of particles measured by $St$. Remarkably in the case where the particles' inertia is not too small, $St^2\gg Fr$,  the dependence of $D_{KY}$ on $St$ drops out,
\begin{eqnarray}&&\!\!\!\!\!\!\!\!\!\!\!\!\!\!\!\!\!\!\!
D_{KY}\!=\!\frac{15\pi \int_0^{\infty}E(k)k dk}{32g}\propto Fr,\ \ St^2\gg Fr.\label{finaldimension} 
\end{eqnarray}
This is the regime of strong gravity where particles sediment through many uncorrelated turbulent vortices of characteristic size $\eta$ during their relaxation (reaction) time $\tau$. Thus the particles react to the sum of actions of many independent vortices rather than to individual vortex. This brings Gaussian statistics that leads to simple relation between rates of particles' volumes contraction (proportional to the average of the fourth power of flow gradients that can be written as square of pair correlations by Gaussianity) and stretching (proportional to pair correlations). The resulting dependence on $\tau$ disappears in the rates' ratio that determines $D_{KY}$. 

Since the spatial distribution of particles is determined uniquely by $D_{KY}$ which in turn is independent of the properties of the particles then we conclude that the particles with $St^2\gg Fr$ distribute in space in universal, size-independent way. 

Furthermore $D_{KY}$ is written in terms of the integral characteristics $\int_0^{\infty}E(k)k dk$ of the turbulence spectrum $E(k)$. This can be interpreted as saying that fluctuations of turbulence with different wave-numbers contribute additively to the fractal formation with the contribution of wave-number $k$ proportional to $E(k)k$. The integral is determined by $k\eta\sim 1$ and can be written in the form 
\begin{equation}
\int_0^{\infty} E(k) k dk=\frac{\epsilon\int_0^{\infty} E(k) k dk}{2\nu\int_0^{\infty} E(k) k^2 dk}=\frac{\epsilon\eta_E}{2\nu}
\end{equation}
where we introduced the ``spectral" viscous scale 
\begin{equation}
\eta_E=\frac{\int_0^{\infty} E(k) k dk}{\int_0^{\infty} E(k) k^2 dk},
\end{equation}
and used $\epsilon=2\nu \int_0^{\infty} E(k) k^2 dk$. Using $\eta_E$ the formula for Kaplan-Yorke dimension takes the form 
\begin{eqnarray}&&\!\!\!\!
D_{KY}=\frac{15\pi \epsilon\eta_E}{64g\nu}.
\end{eqnarray}
Since $\eta_E\sim \eta$ then $\nu/\eta_E$ is the characteristic velocity $v_{\eta}$ of eddies at the Kolmogorov scale. Thus 
\begin{eqnarray}&&\!\!\!\!
D_{KY}\propto\frac{\epsilon}{gv_{\eta}}, \label{parameterdep}
\end{eqnarray}
where the proportionality constant is of order one if intermittency can be neglected. Indeed, the deviation of the scaling exponent of the spectrum from the Kolmogorov's $-5/3$ which is caused by the intermittency is known to be small. To see the relevant quantities we consider the ansatz for the spectrum  
\begin{eqnarray}&&\!\!\!\!
E(k)=C\epsilon^{2/3}k^{-5/3}\exp\left[-\left(\frac{k}{k_{cutoff}}\right)^{\beta}\right],
\end{eqnarray}
where $C$ is the Kolmogorov constant and the exponent $\beta$ characterizes the spectrum's decay in the dissipative range with $k_{cutoff}\eta\sim 1$. This form of the spectrum has to hold at high $k$ only because the integrals over powers of $k$ are determined by $k\sim k_{cutoff}$,
\begin{eqnarray}&&\!\!\!\!
\int_0^{\infty}k^{\delta}E(k)dk=\frac{C\epsilon^{2/3}k_{cutoff}^{\delta-2/3}\Gamma((\delta-2/3)/\beta)}{\beta},
\end{eqnarray}
when $\delta>2/3$ where $\Gamma(x)$ is the Gamma function.
Due to $\epsilon/(2\nu)=\int_0^{\infty}E(k)k^2 dk=C\epsilon^{2/3}\Gamma(4/3\beta)k_{cutoff}^{4/3}/\beta$ the constant $C$ and $k_{cutoff}$ are connected by 
\begin{eqnarray}&&\!\!\!\!
\frac{1}{k_{cutoff}}=\eta \left[\frac{2 C\Gamma(4/3\beta)}{\beta}\right]^{3/4},\nonumber
\end{eqnarray}
where $\eta=(\nu^3/\epsilon)^{1/4}$.
For the integral in the spectral viscous scale we find 
\begin{eqnarray}&&\!\!\!\!
\int_0^{\infty}E(k)k dk=\frac{C\epsilon^{2/3}k_{cutoff}^{1/3}\Gamma(1/3\beta)}{\beta},
\end{eqnarray}
which implies
\begin{eqnarray}&&
\eta_E=\frac{\int_0^{\infty} E(k) k dk}{\int_0^{\infty} E(k) k^2 dk}=\frac{\Gamma(1/3\beta)}{k_{cutoff}\Gamma(4/3\beta)}
\nonumber\\&&
=\left[\frac{2 C\Gamma^{4/3}(1/3\beta)}{\beta \Gamma^{1/3}(4/3\beta)}\right]^{3/4}\eta.
\end{eqnarray}
In the case of exponential decay, $\beta=1$, we find $\eta_E=\eta[2C 3^{1/3}\Gamma(1/3)]^{3/4}\approx 4.6 C^{3/4}\eta$. For Gaussian decay of the spectrum, $\beta=2$, we find 
$\eta_E=\eta[C \Gamma^{4/3}(1/6)/ \Gamma^{1/3}(2/3)]^{3/4}\approx 5.2 C^{3/4}\eta$. Finally in the case of sharp cutoff of the spectrum at $k=k_{cutoff}$, $\beta\to \infty$ we find $\eta_E=4[3C/2]^{3/4}\eta\approx 5.4 C^{3/4}\eta$. Thus $\eta_E/\eta$ is determined by the spectrum's characteristics, $C$ and $\beta$ where $\eta_E/\eta\sim 5$. This agrees well with the experimentally observed $\eta_E/\eta\sim 5.5$. 

We observe from Eq.~(\ref{parameterdep}) that $D_{KY}$ is the inverse ratio of the work $gv_{\eta}$ that gravitational force does on Kolmogovorov scale eddies per unit time per unit mass of the fluid divided by the energy that dissipates into heat at that scale. The studied regime of small $D_{KY}$ is the regime where gravity injects much more energy into the fluid than is dissipated.

It is possible to rewrite $\eta_E$ in terms of the real space statistics of turbulence observing that second order structure function of turbulence $S_2(r)=\langle \left([\bm u(\bm r)-\bm u(0)]\cdot{\hat r}\right)^2\rangle$ obeys (cf. \cite{FH})
\begin{eqnarray}&&
\int_0^{\infty} \frac{S_2(r)dr}{r^2}=\int_0^{\infty} \frac{dr}{r^2}\int [1-({\hat k}\cdot{\hat r})^2]\left(1-\exp[i\bm k\cdot\bm r]\right)\nonumber\\&&\frac{E(k)d\bm k}{2\pi k^2}
=\frac{\pi}{4}\int_0^{\infty} E(k)kdk.\nonumber
\end{eqnarray}
Thus the previous answers can be rewritten in the form 
\begin{eqnarray}&&\!\!\!\!\!\!\!\!\!\!\!\!\!\!\!\!\!\!
\eta_E=\frac{8\nu}{\pi \epsilon}\int_0^{\infty} \frac{S_2(r)dr}{r^2},\ \ 
D_{KY}=\frac{15}{8g}\int_0^{\infty} \frac{S_2(r)dr}{r^2}.
\end{eqnarray}
This form clarifies that $\eta_E$ is the crossover scale between the inertial and viscous ranges of turbulence. Indeed when $r$ is in the viscous range $S_2(r)$ is quadratic in $r$ due to differentiability of the flow, but when $r$ is in the inertial range $S_2(r)$ grows slower than $r$ (in Kolmogorov theory $S_2\propto r^{2/3}$) so that the integral is determined by the crossover between the regimes. We found that in numerical simulations it is simpler to test these real space formulas together with Eq.~(\ref{lambstr}) below than their form in terms of $E(k)$.  

The effect of gravity is most remarkable when $St\gtrsim 1$ where without gravity sling effect dominates the motion \cite{FFS1,Bewley}. The particles move out of turbulent vortices ballistically which produces intersecting streams of particles at the same spatial point with no uniquely defined velocity. Gravity with $Fr\ll 1$ restores the uniqueness by destroying the coherent action of turbulent vortices on particles - the sedimenting particle leaves the vortex before the latter can catch it to produce the sling, cf. \cite{Becgr,Gust}. However gravity influences also the case of $St\ll 1$ which is well studied without gravity where \cite{FHa,FFS1,Fouxon1} 
\begin{eqnarray}&&\!\!\!\!\!\!\!\!\!\!\!\!\!\!\!\!\!\!
D_{KY}\!=\!\frac{\tau^2}{2|\lambda_3|} \int_{-\infty}^{\infty}\!\!\!\! \left\langle \phi[0, 0] \phi[t, \bm x_u(t, 0)] \right\rangle dt,\  St\ll Fr,\label{grneg}
\end{eqnarray}
where $\phi$ is the Laplacian of turbulent pressure, $\lambda_3$ is the third Lyapunov exponent of fluid particles, $\bm x_u(t, 0)$ are the Lagrangian trajectories of the fluid and $St\ll Fr$ is the condition of negligibility of gravity. In this case $D_{KY}$ depends on different time statistics of turbulence. When $St\gtrsim Fr$ gravity changes the form of $D_{KY}$. In the case of $St\gg Fr$ but $St^2\ll Fr$ one has 
\begin{eqnarray}&&\!\!\!\!\!\!\!\!\!\!\!\!\!\!\!\!\!\!
D_{KY}\!=\!\frac{5\tau^2 \int_0^{\infty}\!\! k^3 E_p(k)dk}{2\int_0^{\infty} E(k)kdk}\!\propto\!St^2, \ \ St^2\ll Fr\ll St,\label{prspectrum}
\end{eqnarray} 
where $E_p(k)$ is the spectrum of turbulent pressure. Thus $D_{KY}$ starts to be determined by the instantaneous statistics of turbulence. Both when $Fr\gg St$ and $St^2\ll Fr\ll St$ the Kaplan-Yorke dimension is proportional to $St^2$ but the coefficients of proportionality are different. Since these coefficients depend on high-order moments of velocity gradients for which intermittency is relevant then they depend on Reynolds number. Thus the interplay of gravity and intermittency produces a different dependence on the Reynolds number in properties of fractal distribution of particles in space. 

Finally in the case $St^2\gg Fr$ the formula (\ref{finaldimension}) holds. 

Provided results describe the properties of the spatial distribution of particles in the steady state. Since fluctuations of concentration occur below the Kolmogorov scale then they directly reflect the properties of motion of particles at those scales which are described by the Lyapunov exponents. We determined the Lyapunov exponent. The first Lyapunov exponent obeys 
\begin{eqnarray}&&\!\!\!\!\!\!\!
\lambda_1\tau=\frac{\pi\int_0^{\infty} E(k)kdk}{5 g}\propto Fr,\ \ St\gg Fr,
\end{eqnarray}
that is the product $\lambda_1\tau$ is universal, particles-independent number determined by the spectrum of turbulence. Using the spectral viscous scale we can write
\begin{eqnarray}&&\!\!\!\!\!\!\!
\lambda_1=\frac{\pi\epsilon\eta_E}{10g\tau\nu},\ \ Fr\ll \min[St, 1].
\end{eqnarray}
Finally in terms of the second-order structure function of turbulence we can write 
\begin{eqnarray}&&
\lambda_1\tau=\frac{4}{5g}\int_0^{\infty} \frac{S_2(r)dr}{r^2},\ \ Fr\ll \min[St, 1],\nonumber\\&&
\left|\sum \lambda_i\right|\tau=\frac{3}{2g^2}\left(\int_0^{\infty} \frac{S_2(r)dr}{r^2}\right)^2.
\label{lambstr}
\end{eqnarray}
We observe that $\lambda_1\tau_{\eta}\sim Fr/St$ which is much smaller than one in the considered range. Since the Lyapunov exponent of fluid particles is of order $1/\tau_{\eta}$ then we conclude that gravity decreases the rate of logarithmic divergence of particles. In the remaining asymptotic range of $St\ll Fr$ gravity is negligible so that $\lambda_1$ is equal to the Lyapunov exponent of fluid particles. We note that in contrast to $D_{KY}$ that has three different asymptotic ranges described by Eqs.~(\ref{finaldimension}), (\ref{grneg}), (\ref{prspectrum}), the Lyapunov exponent has only two ranges. To stress the simplicity of the answers we observe that 
\begin{eqnarray}&&\!\!\!\!
\frac{D_{KY}}{\lambda_1 \tau}=\frac{75}{32},\ \ St^2 \gg Fr,\label{results4}
\end{eqnarray}
is universal constant that is independent of properties of particles, turbulence or the strength of gravity. 

Quite similar results hold for $\lambda_3$ that describes the rate of contraction of the smallest dimension of ellipsoid into which turbulence deforms small balls of particles. In the range $St\gg Fr$ one has $\lambda_3\approx -\lambda_1$. When $St\ll Fr$ gravity is negligible and $\lambda_3$ is equal to the third Lyapunov exponent of fluid particles.

The behavior of the sum of Lyapunov exponents $\sum\lambda_i$ that describe the logarithm rate of growth (or rather decrease) of small volumes of particles is readily inferred from the above results by $\sum\lambda_i=D_{KY}\lambda_3$. 

The remaining second Lyapunov exponent describes the rate of growth of areas (given by $\lambda_1+\lambda_2$) or of intermediate axis of the ellipsoid into which turbulence deforms small balls of particles. This has less universal behavior than quantities described above except for one case,
\begin{eqnarray}&&
\lambda_2\tau=-\frac{[\pi \int_0^{\infty}E(k)k dk]^2}{32g^2}\propto Fr^2,\ \ St\gg 1. 
\end{eqnarray}
Thus for strongly inertial particles with $St\gg 1$ the second Lyapunov exponent is negative. This is in contrast to the second Lyapunov exponent of fluid particles $\lambda_2^{turb}$ which is positive. Thus turbulence deforms small balls of inertial particles into cigars and passive tracers into pancakes. When $Fr\ll St\lesssim 1$ there is a correction to the above formula that involves triple correlation function of turbulent gradients and can be found in the main text. That correction which sign is not clear dominates $\lambda_2$ at $Fr\ll St\ll 1$ not allowing to fix the sign of $\lambda_2$. In all cases of $St\gg Fr$ the ratio of $\lambda_2$ to $\lambda_1$ is much smaller than one so that the intermediate axis of the ellipsoid is approximately constant. Finally when $St\ll Fr$ one has $\lambda_2\approx \lambda_2^{turb}>0$ where $\lambda_2^{turb}\tau_{\eta}\sim 1$.

If one can introduce the flow of particles then the particles distribute over fractal in space. We demonstrated that the particles stop to form fractal at the boundary of the asymptotic range of parameters where one can introduce the flow which $Fr\sim 1$, $St\sim 1$, cf. \cite{Becgr}. 

It seems fair to say that no results comparable in scope and detail ever existed in the theory of particles in turbulence including fluid particles themselves. Say $\lambda_1$ for fluid particles is highly non-trivial function of the statistics of the velocity gradients that depends on different time correlations. 

Further we determined numerically the cross-correlations between the locations of particles and of turbulent vortices. These measure quantitatively preferential concentration of particles in strain regions. In the range of parameters where comparison with the theory \cite{Fouxon1} is possible the results confirmed theoretical predictions.


The separation of particles due to white noise $\sigma$ is different in vertical and horizontal directions \cite{FH}. This is likely to produce a difference in the structure of the fractal in horizontal and vertical directions in accord with \cite{Park2014}. The resulting fractal geometry is the topic of the study in progress \cite{FouxonLee}. 

Finally we note that the fractal at the scale $l\ll \eta$ forms at the time scale of order $|\lambda_3|^{-1}\ln (\eta/l)$ which is of order $|\lambda_3|^{-1}$ that we demonstrated to be of order of Kolmogorov time-scale at $St\ll Fr$ and of order $St/Fr$ times the Kolmogorov time-scale otherwise. This time-scale is much smaller than the integral time-scale of turbulence unless $St/Fr$ is unrealistically large. Thus the described phenomena hold for quasi-stationary turbulence as well.

I. F. thanks Roei Harduf for numerous discussions of the flow of particles in the presence of gravity which consideration at $St\ll 1$ started in \cite{FFS} formed the basis of \cite{FHa}. When this work was close to finishing we learnt of the paper \cite{Becgr}. Our results in the questions that were considered in both works are consistent.

This research was supported by a National Research Foundation of Korea (NRF) grant funded by the Korean government (MSIP) (20090093134, 2014R1A2A2A01006544).




\begin{thebibliography}{99}

\bibitem{Frisch} U. Frisch, {\it Turbulence: The Legacy of A. N. Kolmogorov}, Cambridge University Press (1995).

\bibitem{HP} H. G. E. Hentschel and I. Procaccia, Phys. D \textbf{8}, 435 (1983).

\bibitem{KY} J. L. Kaplan and J. A. Yorke, {\it Functional Differential Equations and Approximations of Fixed Points}, Ed. H. O. Peitgen and H. O. Walther, Springer, 204 (1979).

\bibitem{Bec} J. Bec, Phys. Fluids {\bf 15}, L81 (2003).

\bibitem{Becgr} J. Bec, H. Homann, and S. S. Ray, Phys Rev Lett. {\bf 112}, 184501 (2014).

\bibitem{Gust} K. Gustavsson, S. Vajedi, and B. Mehlig, Phys. Rev. Lett. {\bf 112}, 214501 (2014).

\bibitem{Park2014} Y. Park and C. Lee, Phys. Rev. E {\bf 89}, 061004(R) (2014).

\bibitem{FHa}  I. Fouxon and R. Harduf, unpublished; R. Harduf, M. Sc. Thesis, Tel Aviv University. 

\bibitem{Fouxon1} I. Fouxon, Phys. Rev. Lett., {\bf 108}, 134502 (2012). 

\bibitem{Fouxon2} I. Fouxon, arXiv:1110.1625.

\bibitem{Maxey} M. R. Maxey, J. Fluid Mech. {\bf 174}, 441 (1987).

\bibitem{FFS1} G. Falkovich, A. Fouxon and M. Stepanov, Nature {\bf 419}, 151 (2002).

\bibitem{Shaw} R. A. Shaw, Annu. Rev. Fluid Mech., {\bf 35}, 183 (2003).

\bibitem{Seinfeld} J. H. Seinfeld and S. N. Pandis, {\it Atmospheric chemistry and physics: From air pollution to climate change (2nd edition)}, Wiley (2006).

\bibitem{Flagan} R. C. Flagan and J. H. Seinfeld, {\it Fundamentals of air pollution engineering}, Prentice-Hall (1988).

\bibitem{planetary1} A. Bracco, P. H. Chavanis, A. Provenzale and E. A. Spiegel, Phys. Fluids, {\bf 11}, 2280 (1999).

\bibitem{Engineering1} C. T. Crowe, M. Sommerfeld and Y. Tsuji, {\it Multiphase flows with droplets and particles}, CRC (1998).

\bibitem{Engineering2} W. A. Sirignano, {\it Fluid dynamics and transport of droplets and sprays}, Cambridge University Press (1999).

\bibitem{Biology1} G. K{\'a}rolyi, {\'A}. P{\'e}ntek, I. Scheuring, T. T{\'e}l, and Z. Toroczkai, Proc. Natl. Acad. Sci. USA, {\bf 97}, 13661 (2000).

\bibitem{Biology2} T. Nishikawa, Z. Toroczkai, C. Grebogi and T. T{\'e}l, Phys. Rev. E, 65, 026216 (2002).

\bibitem{Nature2013}  W. M. Durham,	E. Climent,	M. Barry, F. De Lillo, G. Boffetta,	M. Cencini, and R. Stocker, Nature Comm. {\bf 4}, 2148 (2013).

\bibitem{FP1} G. Falkovich and A. Pumir, J. Atmos. Sciences {\bf 64}, 4497 (2007). 

\bibitem{FP2} G. Falkovich and A. Pumir, Phys. Fluids, {\bf 16}, L47 (2004). 

\bibitem{MaxeyRiley} M. R. Maxey and J. J. Riley, Phys. Fluids {\bf 26}, 883 (1983).

\bibitem{BFF} E. Balkovsky, G. Falkovich and A. Fouxon, arxiv:chao-dyn/9912027;
Phys. Rev. Lett. {\bf 86}, 2790 (2001).

\bibitem{Falkovich} G. Falkovich and A. Pumir, Phys. Fluids {\bf 16}, L47 (2004).

\bibitem{BGH} J. Bec, K. Gawedzki, and P. Horvai, Phys. Rev. Lett. \textbf{92},
224501 (2004).

\bibitem{Collins} J. Chun, D. L. Koch, S. L. Rani, A. Ahluwalia, and L. R. Collins, J. Fluid Mech. {\bf 536}, 219 (2005).

\bibitem{BecCenciniHillerbranddelta} J. Bec, M. Cencini, and R.
Hillerbrand Phys. Rev. E {\bf 75}, 025301 (2007).

\bibitem{Stefano} J. Bec, L. Biferale, M. Cencini, A. Lanotte, S.
Musacchio, and F. Toschi, Phys. Rev. Lett. {\bf 98}, 084502 (2007).

\bibitem{Cencini} E. Calzavarini, M. Cencini, D. Lohse, and F, Toschi, Phys. Rev. Lett. {\bf 101}, 084504 (2008).

\bibitem{Olla} P. Olla, Phys. Rev. E \textbf{81}, 016305 (2010).

\bibitem{MehligWilkinson} M. Wilkinson, B. Mehlig and K. Gustavsson, Europhys. Lett. {\bf 89}, 50002 (2010).

\bibitem{MW} M. Wilkinson, B. Mehlig, and V. Bezuglyy, Phys. Rev. Lett. {\bf 97}, 048501 (2006).

\bibitem{FFS} G. Falkovich, I. Fouxon, and M. Stepanov, unpublished.

\bibitem{CYL}
J.-I. Choi, K. Yeo, C. Lee, Phys. Fluids {\bf 16}, 779 (2004).

\bibitem{LYC}
C. Lee, K. Yeo, J.-I. Choi, Phys. Rev. Lett. {\bf 92}, 144502 (2004).

\bibitem{JYL}
J. Jung, K. Yeo, C. Lee, Phys. Rev. E {\bf 77}, 016307 (2008).

\bibitem{CKL}
Y. Choi, B.-G. Kim, C. Lee, Phys. Rev. E {\bf 80}, 017301 (2009).

\bibitem{AL1}
A. H. Abdelsamie, C. Lee, Phys. Fluids {\bf 24}, 015106 (2012).

\bibitem{AL2}
A. H. Abdelsamie, C. Lee, Phys. Fluids {\bf 25}, 033303 (2013).

\bibitem{reviewp} G. Falkovich, K. Gawedzki, and M. Vergassola, Rev. Mod. Phys., {\bf 73}, 913 (2001).

\bibitem{Eckmann} J.-P. Eckmann and D. Ruelle, Rev. Mod. Phys. {\bf 57}, 617 (1985).

\bibitem{Dorfman} J. R. Dorfman, {\it An Introduction to Chaos in Nonequilibrium Statistical Mechanics}, Cambridge University Press (1999).

\bibitem{FH} I. Fouxon and P. Horvai, Phys. Rev. Lett. {\bf 100}, 040601 (2008).

\bibitem{review} B. J. Devenish, P. Bartello, J.-L. Brenguier, L. R. Collins, W. W. Grabowski, R. H. A.
IJzermans, S. P. Malinowski, M. W. Reeks, J. C. Vassilicos, L.-P. Wang and Z.Warhaft, Q. J. R. Meteorol. Soc. {\bf 138}, 1401 (2012).

\bibitem{FF} G. Falkovich and A. Fouxon, New J. Phys. {\bf 6}, 50 (2004); G. Falkovich and A. Fouxon, arXiv:nlin.cd/0312033 15 Dec 2003.

\bibitem{Bewley} G. P. Bewley, E. W. Saw, and E. Bodenschatz, New J. Phys., {\bf 15}, 083051 (2013).

\bibitem{Ma} S.-K. Ma, {\it Statistical Mechanics}, World Scientific Publishing Company (1984).

\bibitem{FB}  E.~Balkovsky and A.~Fouxon, Phys. Rev. E {\bf 60}, 4164 (1999).

\bibitem{Batchelor} G. K. Batchelor, {\it An Introduction to Fluid Dynamics}, Cambridge University Press (2000).

\bibitem{Bec2006}
J. Bec, L. Biferale, G. Boffetta, M. Cencini, S. Musacchio and F. Toschi, Phys. Fluids {\bf 18}, 091702 (2006).

\bibitem{GirPope} S. Girimaji and S. Pope, J. Fluid Mech. {\bf 220}, 427 (1990).

\bibitem{Lillo} F. De Lillo, G. Boffetta, and S. Musacchio, Phys. Rev. E {\bf 85}, 036308 (2012).

\bibitem{lectures} I. Fouxon, lecture notes, available online. 

\bibitem{FouxonLee} I. Fouxon and C. Lee, in preparation. 

\bibitem{Maxey2} M. R. Maxey, J. Fluid Mech. {\bf 174}, 441 (2006).

\bibitem{Calza} E. Calzavarini, M. Cencini, D. Lohse, and F. Toschi, Phys. Rev. Lett. {\bf 101}, 084504 (2008).

\end{thebibliography}
\end{document}